%% file: PotykaSchulte2023_arXivV2_ImmiscibleInteractionVOF.tex
 \documentclass[preprint,times,authoryear]{elsarticle_mod_arXiv}
\usepackage{amsmath}
\usepackage{amssymb}
\usepackage{graphicx}
\usepackage{dcolumn}
\usepackage{placeins}
\usepackage{bm}

\usepackage[utf8]{inputenc}
\usepackage[T1]{fontenc}
\usepackage{mathptmx}
\usepackage{etoolbox}
\usepackage{tcolorbox} 
\usepackage{tikz} 
\usepackage{pgfplots} 
\pgfplotsset{width=\linewidth, compat=1.6} 
\usepackage{hyperref}
\usepackage{setspace}



\graphicspath{{pictures/}}

\begin{document}
\begin{frontmatter}

\title{A Volume of Fluid Method for Three Dimensional Direct Numerical Simulations of Immiscible Droplet Collisions}

\author{Johanna Potyka \corref{cor1} \href{https://orcid.org/0000-0003-1310-4434}{\includegraphics[height=10pt]{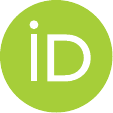}}}%
\author{Kathrin Schulte \href{https://orcid.org/0000-0001-8650-5840}{\includegraphics[height=10pt]{orcid_logo.pdf}}}%

\cortext[cor1]{Email for correspondence: johanna.potyka@itlr.uni-stuttgart.de}%

\affiliation{organization={Institute of Aerospace Thermodynamics (ITLR), University of Stuttgart}, 
    addressline={\mbox{Pfaffenwaldring 31}}, 
    postcode={70569},
    city={Stuttgart}, 
    country={Germany},
    }%
    
\date{\today}
            
%
\begin{abstract}
This paper presents an advanced Volume of Fluid (VOF) method that enables performant three dimensional Direct Numerical Simulations (DNS) of the interaction of two immiscible fluids in a gaseous environment with large topology changes, e.g., binary droplet collisions. %
One of the challenges associated with the introduction of a third immiscible phase into the VOF method is the reconstruction of the phase boundaries near the triple line in arbitrary arrangements. %
For this purpose, an efficient method based on a Piecewise Linear Interface Calculation (PLIC) is shown. %
Moreover, the surface force modeling with the robust Continuous Surface Stress (CSS) model was enhanced to treat such three-phase situations with large topology changes and thin films. %
A consistent scaling of the fluid properties at the interfaces ensures energy conservation. \\
The implementation of these methods in the multi-phase flow solver Free Surface 3D (FS3D) allowed a successful validation. A qualitative comparison of the morphology in binary collisions of immiscible droplets as well as a quantitative comparison regarding the threshold velocities that distinguish different collision regimes shows excellent agreement with experimental results. \\
These simulations enable the evaluation of experimentally inaccessible data like the contributions of kinetic, surface and dissipative energy of both immiscible liquids during the collision process. %
Furthermore, the comparison against binary collisions of the same liquids highlights similarities and differences between immiscible and equal droplet collisions. %
Both can support the modeling of the immiscible liquid interaction in the future.
\end{abstract}
\end{frontmatter}

\input{00_Introduction}

\input{01_Method}

\input{02_Validation}

\input{03_Results}

\input{04_Conclusion}

\section*{Acknowledgment}%
The authors gratefully acknowledge the funding by Deutsche Forschungsgemeinschaft (DFG, German Research Foundation) within the scope of SFB-TRR 75 (project number 84292822) and under  \href{https://www.simtech.uni-stuttgart.de/exc/research/pn/pn1/}{Germany’s Excellence Strategy - EXC 2075 – 390740016}. %
The simulations were conducted on the supercomputer \href{https://www.hlrs.de/solutions/systems/hpe-apollo-hawk}{HPE Apollo (Hawk)} at the \href{https://www.hlrs.de}{High-Performance Computing Center Stuttgart (HLRS)} under the grant no.~FS3D/11142. %
The authors kindly acknowledge the granted resources and continuous support. %

We would like to thank Prof. Carole Planchette and Dr. David Baumgartner from the Technical University of Graz for providing additional experimental reference data and for many fruitful discussions. %
We thank Moritz Heinemann from the University of Stuttgart for the development of the Paraview plugin, which enabled the three dimensional visualization of PLIC interfaces employed in multiple figures in this work. %
Additionally, we would like to thank Prof. Dieter Bothe, Dr.-Ing. Johannes Kromer as well as Prof. Ilia Roisman from the Technical University of Darmstadt for sharing their expertise on multi-phase interface modeling. %

\section*{Data Availability Statement}
Data of the simulation setups and outcomes are openly available at \href{https://doi.org/10.18419/darus-3557}{DOI: 10.18419/darus-3557}. Further data can be provided upon request. %

\section*{Author Declarations}
The Authors have no conflicting interests to disclose.

\section*{CRediT Statement}
\noindent
Johanna Potyka: \\
Conceptualization (equal), Methodology (lead), Software (lead), Validation (lead), Formal analysis (lead), Investigation (lead) , Data Curation (lead), Writing - Original Draft (lead),  Writing - Review \& Editing (support), Visualization (lead)\\
\\
Kathrin Schulte: \\
 Conceptualization (equal), Methodology (support), Software (support), Validation (support), Formal analysis (support), Investigation (support), Data Curation (support), Writing - Original Draft (support),  Writing - Review \& Editing (lead), Visualization (support), Supervision, Project administration, Funding acquisition

\appendix
\input{AppendixLiquidLens}
\input{AppendixEnergy}
\input{AppendixConvergence}
\FloatBarrier

\bibliographystyle{elsarticle-harv}
\bibliography{mybibfile_all}

\end{document}

%% file: 00_Introduction.tex
\section{Introduction} \label{Sec:Introduction}%
The interaction of immiscible liquids is an elementary process in various technical applications ranging from water injection in engines to reduce emissions~\citep{Tsuru2010} to various life science applications. %
In-air microfluidics is a promising approach in this field as a chip-free platform technology that controls collisions of immiscible droplets, for example, to fabricate emulsions and modular 3D biomaterials that have the potential to be used in tissue engineering~\citep{Visser2018, Kamperman2018}. %
Another application of collisions of immiscible droplets is the encapsulation of liquids for pharmaceutical use~\citep{Planchette2010}. %
An enhanced understanding and modeling are crucial for control and further optimization of these processes, which are very complex due to their multi-scale nature. \\

Binary droplet collisions of the same liquid have been studied extensively since the 1990s. %
Various experimental studies concentrated on the transition between different collision regimes (bouncing, coalescence, and separation) and their influencing parameters~\citep{Ashgriz1990, Jiang1992, Qian1997, Orme1997, Pan2019, Gotaas2007, Willis2000, Willis2003} or on detailed investigations of specific regimes~\citep{Pan2008, Pan2009, Roth1999, Roth2007}. \\
Different numerical simulation approaches contributed also to exploring the morphology and outcome of binary droplet collisions of the same liquid. %
\citet{Rieber1995} developed a framework for the interaction of one dispersed liquid in a gaseous environment and studied the outcome of binary droplet collisions with three dimensional Direct Numerical Simulations (DNS). %
They employed the two-phase Volume of Fluid (VOF) method by \citet{Hirt1981} with a Piecewise Linear Interface Calculation (PLIC) by \citet{Youngs1982} for the interface reconstruction. %
\citet{Pan2005} reproduced the experiments by \citet{Ashgriz1990} as well as \citet{Qian1997} with an arbitrary-Lagrangian-Eulerian method combined with a Level-set method. %
They identified the mechanisms of satellite droplet generation for head-on as well as off-center collisions with different impact parameters. %
These findings were confirmed in simulations using a VOF method with a front tracking algorithm to maintain a sharp interface by \citet{Nikolopoulos2009,Nikolopoulos2009a}. %
\citet{Nobari1996} and \citet{Nobari1996a} simulated three dimensional binary droplet collisions with a front tracking method with finite difference discretization. %
By artificially removing the gas film, they enforced temporary or permanent merging of the droplets. %
Thus, they were able to reproduce bouncing, coalescence and separation boundaries experimentally found by \citet{Jiang1992} and \citet{Ashgriz1990} for the head-on collisions. %

Furthermore, different lattice Boltzmann methods (LBM) were employed for two and three dimensional simulations of droplet collisions of the same liquid \citep{Schelkle1995, Inamuro2004, Sakakibara2008, LycettBrown2014, Mazloomi2016, Suzuki2021}. %
\citet{Schelkle1995} proved the applicability of the LBM to the fully three dimensional simulation of single liquid binary droplet collisions. %
\citet{Inamuro2004} investigated the mixing processes of two equal droplets by adding tracer particles to the simulation.  %
\citet{Sakakibara2008} investigated the collision of unequal sized droplets and studied the collision regime boundaries for different diameter ratios. %
The cascaded LBM in three dimensions was validated by \citet{LycettBrown2014} for the simulation of binary drop collisions. %
While \citet{Mazloomi2016} applied an entropic LBM, \citet{Suzuki2021} validated the Lattice Kinetic Scheme (LKS) by reproducing the results of \citet{Pan2019} in different off-center collision regimes. %

While early numerical studies mainly focused on reproducing experiments and validating methods, simulations also provide insights not accessible by experiments. With evaluating the energy balance and dissipation rates, \citet{He2019,He2020} found an explanation for non-monotonic dissipation during bouncing of droplets. %
\citet{Chen2020} investigated bouncing, coalescence and separation from high-resolution DNS with the VOF method. Based on their numerical findings, they deduced an expression for the mass transfer during the collisions. %
Analysing DNS results, \citet{Liu2016} identified the Rayleigh-Plateau instability as the mechanism that determines the rim disintegration in droplet collisions at high initial kinetic energies. %

Various modeling attempts for different binary droplet collisions of the same liquid were made. %
\citet{Ashgriz1990} modeled the outcome of water droplet collisions ignoring dissipation. This is a valid first order model for coalescence and head-on as well as stretching separation of water droplets, but not applicable to more viscous liquids as modeled by \citet{Jiang1992}. %
\citet{Suo2020} proposed corrections and improvements to the latter model. \citet{Munnannur2007} developed a collision model for determining the post-collision velocities and sizes of secondary droplets. %
\citet{Finotello2017} developed a model for the stretching as well as the head-on collision boundary based on a numerical study of droplets of different viscosities. %
A comprehensive overview of those modeling approaches is given by \citet{Sommerfeld2019}.\\
This literature review shows that collisions of the same liquids are already well studied with experiments and numerical simulations and reveals the existence of various models for single-liquid collisions.\\

Collisions of different immiscible liquids, on the other hand, have been studied little so far \citep{Ho2023}. %
Recently, \citet{Ho2023} studied head-on collisions of compound droplets numerically with two-dimensional simulations. %
However, no triple line was present in their simulations. %
The existence of a triple line, however, increases the complexity of experiments, numerical simulation methods and analytical modeling. %
Three interfaces are present now: An interface of liquid~$1$ and liquid~$2$ with interfacial tension $\sigma_{12}$, of liquid~$1$ and gas with surface  tension $\sigma_{13}$ and of liquid~$2$ and gas with surface tension $\sigma_{23}$. %
The relation of the interfacial and surface tensions determines the wetting behavior of the immiscible liquids according to the spreading parameter,
\begin{equation}
S = \sigma_{13}-\sigma_{23}-\sigma_{12}.
\end{equation}
A spreading parameter of $S \geq 0$ indicates full wetting, where one liquid either encapsulates the liquid with the higher surface tension or spreads on it \citep{Planchette2011}. %
Partial wetting is the energetically favorable situation for liquids if $S < 0$.
To our knowledge, there is little data available on the droplet collision of partially wetting liquids.
A small coalescence study by \citet{Wang2004} was incorporated into a larger experimental study focused on the burning characteristics of coalesced droplets of hexadecane and water. %
\citet{ICLASS2021} experimentally described the dewetting phenomenon in the droplet collision of bromonaphtalene with a glycerol solution. %
\citet{Woehrwag2018} presented a LBM for the simulation of drop collisions of immiscible liquids and showed an example of partial drop encapsulation with partial wetting liquids. They focused on low velocity collisions and thus did not simulate separation after the collision.\\
Other studies in the literature focus on fully wetting liquids. %
\citet{Chen2006a} conducted experiments with binary droplet collisions of water and diesel. %
They observed the collision outcomes bouncing, coalescence, single reflex and stretching separation and compared the resulting regime map with the ones of the pure liquids. %
While the bouncing regime is almost identical, they found differences for the separation regimes. %
For example, stretching separation occurs at lower offsets than in collisions of equal droplets. %
\citet{Planchette2009} and \citet{Planchette2010, Planchette2011, Planchette2012} conducted extensive experimental studies with various combinations of glyercol solution and silicon oils focusing on the influence of viscosity on the regime boundary of coalescence, crossing, single-reflex and stretching separation. %
They also identified the mechanism reflexive separation which occurred in the collision of a glycerol solution droplet with a higher density perfloudecaline. %
The aforementioned separation regimes specific to immiscible liquid interaction are explained further in Sec.~\ref{Subsec:ExperimentCompare}.
Based on those experiments, \citet{Roisman2012} provide analytical modeling approaches for major morphological parameters such as the diameter of the collision disc. %
\citet{Planchette2017} extended this work and deduced the fragmentation threshold velocity for the head-on collision of two and three drops of the same liquid and of two drops of immiscible liquids. %
\citet{Baumgartner2022} recently found a universal description for the off-center separation threshold, developed for immiscible droplet-jet interactions as well as single liquid binary droplet collisions. %
The application to binary droplet interaction of immiscible liquids was not shown.

All these modeling approaches are based on energy balances, which depend on some assumptions if derived exclusively from experimental investigations. Numerical simulations can thus make an important contribution to the modeling, as also shown in the work of \citet{Planchette2017} for the collision of equal droplets. %
Furthermore, a universal model requires parameter studies on the influence of the liquids' properties and impact conditions. %
However, broad experimental studies regarding the influence of the liquid properties on the morphology and outcome of collisions of immiscible liquids are difficult, as there is only a limited choice of liquids that can be handled in the laboratory. %
Numerical simulations, in turn, enable extensive parameter studies. \\
\citet{Li2015} presented a Moment of Fluid (MoF) method for the simulation of incompressible flows involving more than two materials. %
Assuming axis symmetry and applying an adaptive mesh refinement close to the contact line, their simulations yielded an adequate resolution and captured the regime transition between coalescence and reflexive separation well for the head-on collisions of diesel and water found from experiments by \citet{Chen2006a}. %
Another numerical study on binary head-on collisions of immiscible droplets with regard to the flow dynamics was presented recently by \citet{Zhang2020}. %
A ternary-fluid diffuse-interface model was applied with an adaptive mesh refinement at the liquid-liquid interface. %
However, their simulations are limited to two dimensions. Furthermore, both liquids have the same density, viscosity, and surface tension, which greatly limits the generality of the study.
A LBM was applied by \citet{Shi2016} for the simulation of off-center bouncing of droplets of immiscible liquids in two dimensions. %
Also \citet{HaghaniHassanAbadi2018} utilized the LBM to simulate head-on bouncing of immiscible droplets in air, again in two dimensions. %
A recent study by \citet{Ebadi2022} employed a phase-field LBM to simulate immiscible droplet interaction in a surrounding liquid (not air) in different collision regimes, but again limited to two-dimensional simulations.\\

As the literature overview shows, there are some studies on the collision of droplets of immiscible liquids, but not to the extent of the droplet collisions of identical liquids. %
Therefore, many questions are still open, e.g., the influence of partial wetting on the separation regime boundaries. %
Furthermore, no comprehensive picture of fully wetting liquids exists, especially concerning the development of the collision complex for asymmetric arrangements, i.e., for off-center collisions, and various combinations of substances. 
Fully three dimensional numerical investigations with high resolution are required here to provide information about the local three dimensional flow field and the triple line movement. \\
The development of a numerical method for the simulation of immiscible liquid interaction in a gaseous environment, its implementation in an existing multi-phase flow simulation program, and its validation was the first objective of the present study. %
The program of our choice is Free Surface 3D (FS3D) \citep{Eisenschmidt2016, Rieber2004}; it employs the VOF method and solves the incompressible Navier-Stokes equations in an one-field formulation by \citet{Hirt1981}. %
FS3D is suitable for performant DNS of multi-phase flows with large topology changes like merging or separation of colliding droplets. %
A Piecewise Linear Interface Calculation (PLIC) by \citet{Youngs1982} is applied to maintain a sharp interface in the two-phase framework. %
This framework is now extended to handle three immiscible phases including triple line motion and thin films.  %
Different models and ideas for two- and three-phase flow available in the literature are combined and extended to achieve this and to overcome various challenges arising from the introduction of a second disperse immiscible liquid to the VOF method. \\%
An interface reconstruction and advection for three immiscible phases were two of those challenges. %
\citet{Benson2002} formulated a great need for methods which enable a topology capturing multi-material reconstruction beyond the layered "onion skin" model. %
Thus, we developed a fast and efficient three-phase PLIC method that reproduces both thin films and junctions of the interfaces at the contact line without prescribing the desired configuration. %
We combine three approaches to an efficient PLIC algorithm: An efficient three-phase positioning algorithm presented by \citet{Kromer2021ArXiv}, an iterative procedure to find the interface orientations proposed in its original form by \citet{Pathak2016}, and a projection method used in multiple studies, e.g., by \citet{Patel2017} and \citet{Washino2010}.\\

The choice of a surface force model suitable for large topology changes poses another challenge. %
The Continuous Surface Stress (CSS) model by \citet{Lafaurie1994} is able to capture large topology changes, but, in its original form, is only applicable to two-phase flow as it does not allow for the computation close to triple lines and thin films. %
A superposition approach according to \citet{Smith2002} as a modification of the CSS model is the solution in the present work. %
Furthermore, balancing the surface forces in the CSS model following the principle of the balanced Continous Surface Force (bCSF) approach by \citet{Popinet2009} lead to energy conserving simulations. \\

A second aim of this work was an application of the framework to obtain the energy contributions for exemplary interactions of immiscible liquids, namely binary droplet collisions. %
The energy budget can be obtained in such detail only from numerical simulations and, together with geometrical considerations, forms the basis for modeling the collision processes. %
Another important contribution towards the modeling is provided by the comparison with collisions of the same liquids, which is also shown in this study. %
This makes it possible to identify similarities and differences and thus possibilities for extending existing single-fluid models. The modeling, however, is beyond the scope of our present work.

The paper is organized as follows:  Section~\ref{Sec:Method} presents the methods chosen for the extension of the VOF method to two immiscible liquids interacting in air as well as their necessary adaptations. %
Section~\ref{Sec:Validation} shows a validation of the implemented three-phase VOF method with an analytic test case and experimental results. %
The evaluation of the energy balance for different binary droplet collision regimes and a comparison to single liquid collisions is presented in Sec.~\ref{Sec:Results}. %

%% file: 01_Method.tex
\section{Method} \label{Sec:Method}%
This section describes the main governing equations, the applied VOF method with its required extensions as well as the software used in this study. %
The term "three-phase" indicates an immiscible liquid-liquid-gas situation throughout this study. %
However, the method is equally applicable to systems of three liquids. %
The indices $m,n \in \{1,2,3\}$ are used to identify the phases, 1: inner, encapsulated liquid, 2: outer, encapsulating liquid and 3: gas phase.\\
The one-field formulation of the incompressible Navier-Stokes equations is employed to describe the interaction of the two immiscible liquids in air. The momentum conservation
\begin{equation}
  \rho \left[\frac{\partial\mathbf{u}}{\partial t} + \mathbf{\nabla}\cdot \left(\mathbf{u}\mathbf{u}\right)\right] = -\mathbf{\nabla} p + \mathbf{\nabla} \cdot \mathbf{S}+ \mathbf{f}_{\Gamma} 
  \label{Eq:Momentumconservation} %
\end{equation} %
is solved along with the mass %
\begin{equation}
 \frac{\partial \rho}{\partial t} + \mathbf{\nabla}\left(\rho \mathbf{u}\right) = 0
\end{equation}
and volume conservation %
\begin{equation}
  \mathbf{\nabla} \cdot \mathbf{u} = 0\label{Eq:Volumeconservation}
\end{equation} %
with $\mathbf{u}$ the  velocity, $\rho$ and $\mu$ the volume weighted cell averaged density and dynamic viscosity, $p$ the pressure,  and $\mathbf{f}_{\Gamma}$ the interfacial forces due to surface and interfacial tensions  $\sigma_{mn}$ of phase $m$ and $n$. %
Gravity is not considered in the present study. %
Additionally, isothermal conditions without phase change are assumed. %
Thus the energy equation can be neglected for the solution of this pure hydrodynamics problem. %
For each of the phases constant material properties $\rho_m$, $\mu_m$ and $\sigma_{mn}$ are prescribed within this work. %

Since Newtonian fluids are assumed, %
\begin{equation}
  \mathbf{S} = \mu  \left[\mathbf{\nabla} \mathbf{u}+\left(\mathbf{\nabla} \mathbf{u}\right)^T\right]
\end{equation} %
describes the viscous stress tensor. The surface forces $\mathbf{f}_{\Gamma}$ only act at an interface of two different phases. %
The formulation of these interfacial forces as a volume force acting in cells of the discrete grid which are close to the interface is discussed in Sec.~\ref{Subsec:SurfaceForces}. %
The evolution of the velocity $\mathbf{u}$ over time is given by the conservation laws according to Eq.~\ref{Eq:Momentumconservation} and Eq.~\ref{Eq:Volumeconservation} together with the initial and boundary conditions. %
All terms of Eq.~\ref{Eq:Momentumconservation} are evaluated subsequently.  %
As an incompressible flow is considered, the pressure gradient $\nabla p$ results from the elliptical Pressure Poisson equation, %
\begin{equation}
\nabla \cdot \left( \frac{1}{\rho} \nabla p \right) = \nabla \cdot \left(  -\left( \mathbf{u} \cdot \nabla \right) \mathbf{u} + \frac{1}{\rho} \left( \nabla \cdot \mathbf{S} + \mathbf{f}_\Gamma \right)\right) \text{.} 
\label{Eq:Pressurepoisson}
\end{equation}
A multi-grid solver with an efficient implementation of the red-black Gauss-Seidel smoother \citep{Wauligmann2021}  was recently implemented for the iterative solution of Eq.~\ref{Eq:Pressurepoisson} in FS3D \citep{HLRSBericht2022Preprint}.  %

The utilized in-house software FS3D \citep{Eisenschmidt2016} employs a Cartesian Marker and Cell grid \citep{Harlow1965}, also called staggered grid, which is equidistant in this study with an edge length $\Delta x$ in all three dimensions. %
In the following, the indices $i$,$j$ and $k$ for the \mbox{$x$-,} $y$- and $z$- direction indicate the index position of the scalar fields such as the pressure or the volume fractions. The vector components are stored at the cell faces.
Further details on the discretization are given in \citet{Eisenschmidt2016} and \citet{Rieber2004}. %
One major advantage of the Cartesian grid lies in the high efficiency achievable with domain decomposition for the parallelization of the software  using Message Passing Interface (MPI). %
 \\
The major changes required to proceed from a three dimensional two-phase VOF Method with a PLIC reconstruction to three immiscible phases with large topology changes is discussed in subsections \ref{Subsec:VOFPLIC} and \ref{Subsec:SurfaceForces}. %

\subsection{Phase and Interface Detection, Reconstruction and Transport with Sharp Interfaces} \label{Subsec:VOFPLIC}%
\begin{figure}[!tb]
\centering
  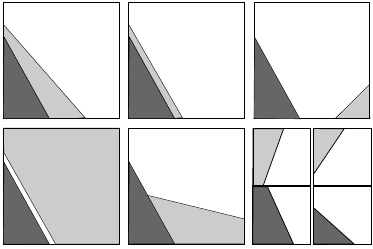
   \caption{Possible three-phase configurations of liquid~1 (dark grey), liquid~2 (light grey) and gas with ((e) and (f)) and without a triple line. %
   The three-phase cells are marked in cyan. Cells with a liquid~2 interface that lie inside the reconstruction stencil of liquid~1 are marked in magenta. %
   }
   \label{Fig:3phTopology}
\end{figure}

\begin{figure}[!tb]
\centering
\includegraphics[scale=1.2]{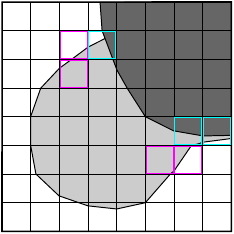}
\caption{Exemplary distribution of two immiscible VOF variables (dark and light grey) in air (white) and their desired interface reconstruction. %
The three-phase cells are marked in cyan, the cells with a liquid~2 interface within the reconstruction stencil of liquid~$1$ are marked in magenta.}\label{Fig:ImmiscibleExample}
\end{figure}

\begin{figure}[!tb]
\centering
 \def\svgwidth{10cm}
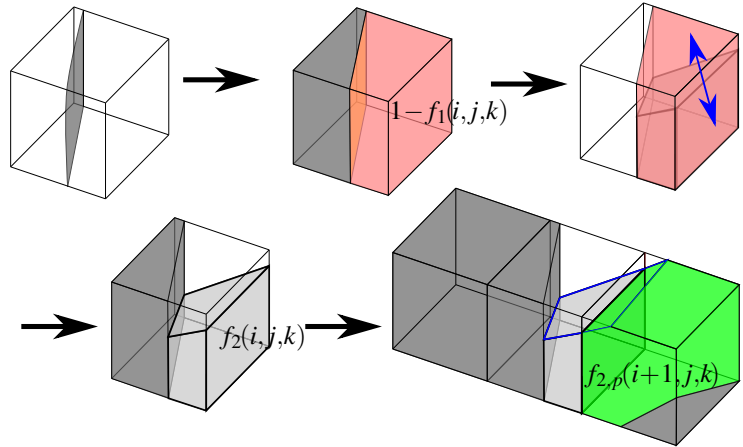
\caption{The position of the liquid~2's (light grey) PLIC interface is found inside the residual polyhedron (red) after cutting the cell with the liquid~1's (dark grey) interface. %
The predicted volume fractions $f_{2,p}$ (green) result from an extension of the plane with the normal $\mathbf{n}_{2}\left(\theta\right)$ and signed distance $l_{2}$ to all neighboring cells. %
The predicted volume fraction $f_{2,p}$ is calculated from the volume enclosed by the extended plane and the volume of the neighboring cell that is not occupied by liquid 1. %
This is depicted for two of the 26 neighboring cells in the $3\times3\times3$-stencil here.}%
\label{Fig:fp}
\end{figure}

The first requirement of the simulation of two immiscible liquids in a gaseous environment is the ability to identify the distribution of each phase. %
A well-known approach to solve the one-field formulation of the Navier-Stokes equations for multi-phase simulations is the VOF method by \citet{Hirt1981}. %
This method allows for solving one single set of conservation equations by introducing the volume fraction 
\begin{equation}
    f_m\left(i,j,k\right) = \frac{V_{m}\left(i,j,k\right)}{V_\mathrm{cell}} =
      \begin{cases}  
         1$, cell with phase~$m$ only$\\ 
         \left]0,1\right[$, cell with interface of~$m \\ 
         0 $, cell without phase~$m \\
      \end{cases}      
      \label{Eq:VOFVar}
\end{equation}
to identify a phase~$m$ with volume $V_m$ inside a cuboidal cell $\left(i,j,k\right)$ of volume $V_{\mathrm{cell}} = \left(\Delta x\right)^3$. %
As the sum of all volume fractions has to be one, two fields of volume fractions for the two liquids ($f_1$ for phase~1 and $f_2$ for phase~2) are sufficient to identify all phases. Possible configurations of cells with three phases where triple lines and thin films can be present and exemplary distributions of two immiscible VOF variables are depicted in Figs.~\ref{Fig:3phTopology} and \ref{Fig:ImmiscibleExample}. The cell-averaged fluid properties within cell $(i,j,k)$ of the Cartesian grid are the volume-weighted averages
\begin{equation}
 \Psi(i,j,k) = \sum_{m=1}^{3} f_m^{N_\text{smooth}+1}(i,j,k) \Psi_m  \text{ with } \Psi \in \{ \rho, \mu \}\label{Eq:rho}
\end{equation}
employing a smoothed volume fraction $f_m^{N_\text{smooth}+1}$. %
The necessity and computational details of this smoothing are discussed in Sec.~\ref{Subsec:SurfaceForces}.\\
The transport equation
\begin{equation}
\frac{\partial f_m}{\partial t} + \mathbf{\nabla} \cdot \left(f_m\mathbf{u}\right) = 0
\label{Eq:fadvect}
\end{equation}
describes the advection of each volume fraction field, $f_m$. %
The velocity $\mathbf{u}$ results from the momentum balance, Eq.~\ref{Eq:Momentumconservation}, in each time-step. %
Thus, the velocity of all three phases is identical inside one cell. 
The split advection method by \citet{Strang1968} is used to solve Eq.~\ref{Eq:fadvect}. %
Each of the three advection steps ($x$-, $y$- and $z$-direction) requires a previous reconstruction of the interfaces to avoid numerical diffusion of the volume fractions $f_m$. %
The PLIC method by \citet{Youngs1982} is employed for this purpose in two-phase cells and adapted for cells with three phases. %
The PLIC plane represents the interface, which truncates the given volume fraction $f(i,j,k)$ from the cell $(i,j,k)$. %
It is defined by its normal $\mathbf{n}_m$ and the signed distance $l_m$ from a local origin within a cell. %
Prior to the interfaces' normal computation (step 2) and the positioning of the interfaces (step 3), a material ordering is the first step (step 1) during the PLIC reconstruction in three-phase cells (it is not required in two-phase cells). %
The liquid with the lower surface tension $\sigma_{23}$ spreads on the other liquid.  %
With the assumption $\sigma_{13} > \sigma_{23}$ valid from now on, phase $2$ always covers phase $1$ and both phases are surrounded by the gaseous continuous phase $3$. %
The second and third steps are relatively easy to perform explicitly for two-phase cells in a Cartesian grid. %
In two-phase cells, the normal of a reconstructed interface 
\begin{equation}
\mathbf{n}_m = \frac{\mathbf{\nabla} f_m}{|\mathbf{\nabla} f_m |} \label{Eq:GradientNormal}
\end{equation}
and the signed distance, $l_m$, of the interface are computed from the volume fraction, $f_m$, according to \citet{Youngs1982} and \citet{Rider1998}. %
This is applicable if only one kind of interface is present inside the $3\times3\times3$ stencil. %
This stencil is necessary for a finite difference discretization of Eq.~\ref{Eq:GradientNormal}. %
The volume fraction distribution of the spreading liquid~$2$ is truncated by the presence of liquid~$1$ in three-phase situations with a triple line, cf. Fig.~\ref{Fig:ImmiscibleExample}. %
Thus, the gradient of the volume fraction no longer yields the orientation of the interface of liquid~$2$. %
The solution employed in our study is a sequential reconstruction (also called nested reconstruction), which is a compromise of accuracy and a performant algorithm \citep{Pathak2016, Kromer2021ArXiv}. %
It combines two three-phase PLIC approaches: an explicit method widely used by, e.g., \citet{Patel2017, Washino2010} and \citet{Sussman2001}, which imposes a prescribed contact angle, and an iterative approach presented by \citet{Pathak2016}. %
The encapsulated liquid $1$ is reconstructed first and independently by the two-phase algorithm. %
An iterative algorithm is used to compute the second interface of the encapsulating liquid $2$ within the residual part of the cell.
 Equation~\ref{Eq:GradientNormal} does not reflect the correct interface orientation due to the truncation of the volume fraction distribution  liquid~$2$ by liquid~$1$. %
However, in the plane parallel to the PLIC plane of the inner liquid, the gradient of $f_2$ has the correct orientation. %
Thus, the orientation of liquid 2's normal in that parallel plane can be obtained by the projection of the gradient onto that plane,
\begin{equation}
 \mathbf{t}_2 = \frac{\mathbf{\nabla} f_{2} - (\mathbf{n}_{1}\cdot \mathbf{\nabla} f_{2}) \mathbf{n}_{1}}{|| \mathbf{\nabla} f_{2} - (\mathbf{n}_{1}\cdot \mathbf{\nabla} f_2) \mathbf{n}_{1}||}\text{.} \label{Eq:projection}
\end{equation}
This projection is used in the aforementioned explicit method \citep{Patel2017,Washino2010}. Thus, the orientation of the second PLIC plane can be found with an iterative method with the aid of the parametrization
\begin{equation}
\mathbf{n}_{2} = \mathbf{n}_{1} \cos(\theta) + \mathbf{t}_2 \sin(\theta) \label{Eq:n_t_theta}
\end{equation}
where the normal of the first PLIC plane $\mathbf{n}_{1}$ results from Eq.~\ref{Eq:GradientNormal}. 
The angle $\theta$ of the second PLIC interface towards the first PLIC interface corresponds to the liquid-liquid contact angle, if a triple line is present in the three-phase cell. %
The optimal choice of $\theta$,  which represents the interface best, is found by a minimization of %
  \begin{equation}
  \begin{aligned}
      g(i,j,k) \left(\theta\right) = & \\
       \sum\limits_{i^*=i-1}^{i+1}\sum\limits_{j^*=j-1}^{j+1} & \sum\limits_{k^*=k-1}^{k+1}   (  f_2 (i^*,j^*,k^*) 
        -  f_{2,p} (i^*,j^*,k^*, \theta))^2 \text{,}\label{Eq:gerr}
       \end{aligned}
  \end{equation}
which is the $L_2$-error of the true volume fraction of the liquid~$2$ compared to a predicted volume fraction $f_{2,p}$. \citet{Pathak2016} employ a similar minimization, but in two angles and do not exploit the more efficient projection in Eq.~\ref{Eq:projection}. %
The predicted volume fraction, $f_{2,p}$ results from extending the PLIC interface of the second liquid with an arbitrary angle $\theta$ into the neighboring cells in a $3\times3\times3$ stencil. %
Analogous to the three-phase PLIC method by \citet{Pathak2016}, the cell in a $3\times3\times3$ stencil with maximum $f_1 f_2 \left(1-f_1-f_2 \right)$ is employed as center cell of the extension. The predicted volume truncated from each neighboring cell within the $3\times3\times3$ stencil is computed and $f_{2,p}$ is calculated. %
An example is illustrated in Fig.~\ref{Fig:fp}. It should be noted, that the extension inside liquid~$1$ is always set to zero. %
During the minimization procedure multiple positionings of the PLIC plane of liquid 2 with multiple volume computations of truncated polyhedrons are necessary. %
This is performed efficiently with the sequential PLIC positioning described by \citet{Kromer2021ArXiv}. %
The efficiency of this positioning algorithm is crucial to the overall performance. %
The minimization of Eq.~\ref{Eq:gerr} is performed with Brent's algorithm like suggested by \citet{Pathak2016}. %
\begin{figure}[!tb]
\centering
 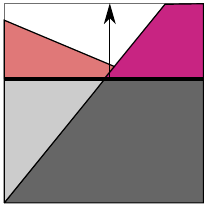
 \caption{Simplified two-dimensional visualization of the polyhedrons necessary for the advection of the volume fractions. %
 }\label{Fig:advect}
\end{figure}

After the successful positioning of the interfaces, the volume fractions are advected according to Eq.~\ref{Eq:fadvect} subsequently in all spatial directions using the splitting by \citet{Strang1968}. %
The advection in three-phase cells again requires an intersection of the reconstructed polyhedron of each phase with a plane. %
Here the plane is parallel to a cell face of the advection direction. %
Figure~\ref{Fig:advect} illustrates this backward flux calculation method to obtain the advected amount of each liquid to the neighboring cell. %

Summarizing, the accurate and efficient PLIC reconstruction and advection close to triple lines is one key element for extending an efficient and highly parallelized two-phase VOF flow solver with reconstructed interfaces to three immiscible liquid and gas phases forming triple lines and thin films. %
Employing the efficient sequential three-phase PLIC positioning algorithm by \citet{Kromer2021ArXiv} together with the described efficient normals' computation with an iteration in one degree of freedom only is crucial for the performance of the parallelized simulation program FS3D. %
\subsection{Surface forces for three-phase situations} \label{Subsec:SurfaceForces}%
\begin{figure}[!tb]
\centering
 \includegraphics[scale=1]{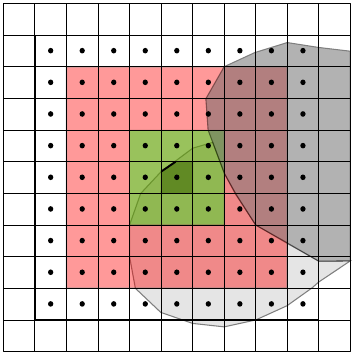}
 \caption{Exemplary visualization of the cells contributing to the computation of the interfacial force, $f_\Gamma$, in the central cell, according to Eq.~\ref{Eq:Superposition} with use of Eq.~\ref{Eq:T}. These cells are marked by black dots. The smoothing of the volume fraction, $f_m^{N_\text{smooth}}$, according to Eq.~\ref{Eq:Smoothing} results in a $3\times3\times3$-stencil (green) if $n_\text{sm.} = 1$, three subsequent smoothing steps, $N_\text{smooth}$ = 3, result in a $7\times7\times7$ stencil (red). As the surface stress tensor, $\mathbf{T}_m$ in Eq.~\ref{Eq:T}, is computed from central differences using  $f_m^{N_\text{smooth}}$, an additional cell layer influences the computation of $f_\Gamma$. Thus, an additional smoothing step, $N_\text{smooth}+1$, is employed for evaluating the density, $\rho^*$ (Eq.~\ref{Eq:densityscaleing}), that scales $f_\Gamma$ in the computation of the acceleration in Eq.~\ref{Eq:Scaling}.
 }\label{Fig:VisSmoothing}
\end{figure}

The interfacial forces, $\mathbf{f}_\Gamma$, in the momentum balance, Eq.~\ref{Eq:Momentumconservation}, that result from the surface tensions, $\sigma_{mn}$, are approximated with the Continuous Surface Stress model (CSS) by \citet{Lafaurie1994}, which is chosen for this work as it is suitable for the simulation of the disintegration of the droplets after a collision at higher velocities. %
If three-phase situations occur, the original CSS model requires enhancement. According to \citet{Smith2002}, the interfacial tensions, $\sigma_{mn}$, can be decomposed into phase-specific partial surface tensions $\gamma_m$ by
\begin{equation}
\sigma_{mn} = \gamma_m + \gamma_n \text{.}
\label{Eq:SigmaDecompose}
\end{equation}
This approach was also applied in other studies \citep{Zhang2019, Joubert2020}.
Equation~\ref{Eq:SigmaDecompose} provides three equations for our three surface- and interfacial tensions. %
Note, that this decomposition cannot be extended beyond three phases as described here. %
The curvature information is obtained from the geometry surface stress tensor \citep{Zhang2019}
\begin{equation}
\mathbf{T}_m = \mathbf{\nabla} \cdot \left(|\nabla f_m^{N_\text{smooth}}| \left(\mathbf{I}-\frac{\nabla f_m^{N_\text{smooth}}}{|\nabla f_m^{N_\text{smooth}}|} \otimes \frac{\nabla f_m^{N_\text{smooth}}}{|\nabla f_m^{N_\text{smooth}}|}\right)\right)
\label{Eq:T}
\end{equation}
for each of the three phases $m$. %
All spatial gradients are approximated with central differences. %
The smoothing of each volume fraction, $f_m$, is required for the application of the CSS model in order to reduce the influence of the grid structure \citep{Brackbill1992}. %
The discretization of the Cartesian grid is otherwise imposed on the gradient computation as the result depends on the fill level of the cell (rather empty or full). %
The smoothed volume fraction, $f_m^{N_\text{smooth}}$, is obtained from applying a linear smoothing in a $3\times3\times 3$ stencil $N_\mathrm{smooth}$-times to the volume fraction $f_m$ in each cell $(i,j,k)$ according to
\begin{equation}
\begin{aligned}
f_m^{n_\mathrm{sm}}(i,j,k) = & \\
\sum_{k_l =-1}^{1} \sum_{j_l =-1}^{1} & \sum_{i_l =-1}^{1} \mathbf{P}\left(i_l,j_l,k_l\right) f_m^{n_\mathrm{sm}-1}(i+i_l,j+j_l,k+k_l) \\
\end{aligned}
 \label{Eq:Smoothing}
\end{equation}
with the linear smoothing point operator
\begin{equation}
\mathbf{P} =
\begin{cases}  
\left( 
\begin{array}{rrr}
1 & 2 & 1 \\ 
2 & 4 & 2 \\
1 & 2 & 1 \\ 
\end{array}
\right) \text{, if } k_l \in \{-1,1\}  \\
\\
 \left( 
\begin{array}{rrr}
2 & 4 & 2 \\ 
4 & 8 & 4 \\
2 & 4 & 2 \\ 
\end{array}
\right) \text{, if } k_l=0 \\
\end{cases}
\end{equation}
and $n_\mathrm{sm} \in \{1,2,..., N_\mathrm{smooth}\}$. In the first smoothing step, $f_m^{n_\mathrm{sm}=1}(i,j,k)$, $f_m^{0}(i,j,k)=f_m(i,j,k)$ inside the whole computational domain. The stencil of the first smoothing step is visualized in green for an exemplary cell in Fig.~\ref{Fig:VisSmoothing}. The resulting stencil after $N_\text{smooth}=3$ smoothing steps is visualized in red in the same figure.

The slightly different quadratic B-spline smoothing proposed by \citet{Brackbill1992} as well as by \citet{Lafaurie1994} leads to an introduction of additional gradients to the computed surface forces and thus to the velocity computation due to the parabolic (instead of linear) decrease of the volume fraction.
This introduces an energy gain.
The here presented smoothing on the other hand is avoiding additional unphysical velocity gradients and energy conserving. %
\citet{Brackbill1992} saw no effect of more than two smoothing steps for the resolutions they could realize in the 1990s. %
However, for the  resolutions of the simulations presented in Sec.~\ref{Sec:Validation}, three to four smoothing steps guarantee energy conservation without introducing unnecessary additional numerical dissipation. A study on the influence of the number of smoothing steps on the energy conservation throughout a droplet collision test case is presented in \ref{App:ConvergenceEnergy}. \\
Now, the interfacial forces can be computed by the superposition
\begin{equation}
\mathbf{f}_\Gamma = \sum\limits_{m=1}^{3}\gamma_m \mathbf{T}_m \text{.}
\label{Eq:Superposition}
\end{equation}
The computed forces are equal to the original two-phase CSS model, if only two phases occur inside the smoothing stencil. %
The superposition in Eq.~\ref{Eq:Superposition} with the decomposed interfacial tensions from Eq.~\ref{Eq:SigmaDecompose} is reasonable, as the forces resulting from the interfacial tensions balance for partially wetting liquids in a stationary state. This is shown in Sec.~\ref{Subsec:SurfaceTest}. The surface forces should drive the liquid across the encapsulated phase, and a stationary state is reached at full encapsulation for fully wetting liquid combinations. %
Note that the superposition is not applicable for the interaction with a solid surface, as the solid cannot deform. %

In which way the acceleration due to the interfacial forces
\begin{equation}
\left(\frac{\partial \mathbf{u}}{\partial t}\right)_\Gamma =\frac{1}{\rho^*} \mathbf{f}_\Gamma
\label{Eq:Scaling}
\end{equation}
is calculated in three-phase situations is often not discussed in the literature on sharp interface VOF methods. %
For two-phase flows, \citet{Popinet2009} emphasizes the necessity of ensuring a balance of the surface and pressure forces in order to minimize spurious currents. %
Choosing the CSF model with an accurate computation of the curvature to obtain a consistent discretization in all terms of the momentum balance, he suggested a scaling of the surface forces with a volume weighted averaging of the density.
This is a reasonable choice for the surface force computation based on the sharp volume fractions he applied. %
However, the CSS model that is applied in the present work in order to capture large topology changes requires the smoothing of the volume fractions as discussed before. %
A scaling of the interfacial forces that are obtained from the smoothed volume fractions with the sharp density field leads to high spurious currents and an unphysical deformation of the liquid phase along with an unphysical momentum and energy gain due to the applied stencil. In cells within the smoothing stencil of the sharp interface, the smeared interfacial forces are multiplied with the low gas density. %
Also, the use of the smoothed volume fractions, $f_m^{N_\text{smooth}}$, for the computation of the density in Eq.~\ref{Eq:Scaling} would take into account cells containing only the density of the gas phase due to the discretization of the volume fractions' gradient in Eq.~\ref{Eq:T} with central differences. %
Therefore, a further smoothing step of the volume fractions, $f_m^{N_\text{smooth}+1}$, is performed for the density calculation, according to
\begin{equation}
 \rho^* = \sum_{m=1}^3  f_m^{N_\text{smooth}+1} \rho_m \text{.}
 \label{Eq:densityscaleing}
\end{equation}
This additional smoothing step results in another layer of cells being included in the smoothing process, and the interfacial forces $\mathbf{f}_\Gamma$ from Eq.~\ref{Eq:Superposition} are now scaled with appropriate densities. %
The reasoning for the additional smoothing step is illustrated in Fig.~\ref{Fig:VisSmoothing} with the values contributing to the center cell's $\mathbf{f}_\Gamma$ marked with \textcolor{black}{\textbullet{}} dots. %
This comes with the drawback, that the interfacial forces are calculated form a slightly different volume fraction distribution $f_m^{N_\text{smooth}}$ than the weighted density $\rho^*(f_m^{N_\text{smooth}+1})$, but the error for the energy conservation is smaller than scaling the forces in a whole layer of cells with a large error in the density, which would result from the several orders of magnitude difference of the densities of the gas and the liquids. %

In order to balance the forces and minimize numerical errors, it is essential to scale the pressure forces accordingly in order to allow a physical balance of forces for a flow at rest \citep{Popinet2009}.
The same is true for the advection as well as the viscous force term in the momentum equation.
Consequently, also the viscosity and density is computed from the smoothed volume fraction field, $f_m^{N_\mathrm{smooth}+1}$, while the VOF advection, cf.\ Eq.~\ref{Eq:fadvect}, is performed with the sharp volume fraction. %
Thus, the smearing of the interface is limited to the same stencil in each time-step, while a sharp interface is transported. %
\\
The superposition of the interfacial forces, Eq.~\ref{Eq:Superposition}, with the aid of a decomposition of the interfacial tensions, $\sigma_{mn}$, into partial interfacial tensions, $\gamma_m$, according to Eq.~\ref{Eq:SigmaDecompose} along with a reasonable choice of the scaling density, $\rho^*$, (Eq.~\ref{Eq:densityscaleing}) and the arising consequences for all other computations of volume weighted liquid properties proved to be one of the main challenges that we overcame in extending a two-phase VOF method to three deformable phases without introducing unphysical energy gain. %
The advantage of the here presented balanced CSS approach is its robustness without requiring an elaborate mesh refinement. 
This makes it easy to apply to flows with thin films, triple lines, air entrapment and large topology changes and well suitable for the efficient parallelization with a simple Cartesian domain decomposition. %

%% file: pictures/DreiphasenKonfigurationen.pdf_tex
\begingroup%
  \makeatletter%
  \providecommand\color[2][]{%
    \errmessage{(Inkscape) Color is used for the text in Inkscape, but the package 'color.sty' is not loaded}%
    \renewcommand\color[2][]{}%
  }%
  \providecommand\transparent[1]{%
    \errmessage{(Inkscape) Transparency is used (non-zero) for the text in Inkscape, but the package 'transparent.sty' is not loaded}%
    \renewcommand\transparent[1]{}%
  }%
  \providecommand\rotatebox[2]{#2}%
  \newcommand*\fsize{\dimexpr\f@size pt\relax}%
  \newcommand*\lineheight[1]{\fontsize{\fsize}{#1\fsize}\selectfont}%
  \ifx\svgwidth\undefined%
    \setlength{\unitlength}{178.58268257bp}%
    \ifx\svgscale\undefined%
      \relax%
    \else%
      \setlength{\unitlength}{\unitlength * \real{\svgscale}}%
    \fi%
  \else%
    \setlength{\unitlength}{\svgwidth}%
  \fi%
  \global\let\svgwidth\undefined%
  \global\let\svgscale\undefined%
  \makeatother%
  \begin{picture}(1,0.66666662)%
    \lineheight{1}%
    \setlength\tabcolsep{0pt}%
    \put(0,0){\includegraphics[width=\unitlength,page=1]{DreiphasenKonfigurationen.pdf}}%
    \put(0.23244621,0.59362957){\color[rgb]{0,0,0}\makebox(0,0)[lt]{\lineheight{1.25}\smash{\begin{tabular}[t]{l}(a)\end{tabular}}}}%
    \put(0.56042485,0.6016615){\color[rgb]{0,0,0}\makebox(0,0)[lt]{\lineheight{1.25}\smash{\begin{tabular}[t]{l}(b)\end{tabular}}}}%
    \put(0.90582751,0.59696204){\color[rgb]{0,0,0}\makebox(0,0)[lt]{\lineheight{1.25}\smash{\begin{tabular}[t]{l}(c)\end{tabular}}}}%
    \put(0.22316493,0.25585788){\color[rgb]{0,0,0}\makebox(0,0)[lt]{\lineheight{1.25}\smash{\begin{tabular}[t]{l}(d)\end{tabular}}}}%
    \put(0.56647193,0.25234496){\color[rgb]{0,0,0}\makebox(0,0)[lt]{\lineheight{1.25}\smash{\begin{tabular}[t]{l}(e)\end{tabular}}}}%
    \put(0.76213695,0.2617158){\color[rgb]{0,0,0}\makebox(0,0)[lt]{\lineheight{1.25}\smash{\begin{tabular}[t]{l}(f)\end{tabular}}}}%
    \put(0.91254247,0.2617158){\color[rgb]{0,0,0}\makebox(0,0)[lt]{\lineheight{1.25}\smash{\begin{tabular}[t]{l}(g)\end{tabular}}}}%
    \put(0,0){\includegraphics[width=\unitlength,page=2]{DreiphasenKonfigurationen.pdf}}%
  \end{picture}%
\endgroup%

%% file: pictures/g_error_pt_Polyedervisualisierung2.pdf_tex
\begingroup%
  \makeatletter%
  \providecommand\color[2][]{%
    \errmessage{(Inkscape) Color is used for the text in Inkscape, but the package 'color.sty' is not loaded}%
    \renewcommand\color[2][]{}%
  }%
  \providecommand\transparent[1]{%
    \errmessage{(Inkscape) Transparency is used (non-zero) for the text in Inkscape, but the package 'transparent.sty' is not loaded}%
    \renewcommand\transparent[1]{}%
  }%
  \providecommand\rotatebox[2]{#2}%
  \ifx\svgwidth\undefined%
    \setlength{\unitlength}{317.87795001bp}%
    \ifx\svgscale\undefined%
      \relax%
    \else%
      \setlength{\unitlength}{\unitlength * \real{\svgscale}}%
    \fi%
  \else%
    \setlength{\unitlength}{\svgwidth}%
  \fi%
  \global\let\svgwidth\undefined%
  \global\let\svgscale\undefined%
  \makeatother%
  \begin{picture}(1,0.6)%
    \put(0,0){\includegraphics[width=\unitlength,page=1]{g_error_pt_Polyedervisualisierung2.pdf}}%
    \put(0.3,0.15){$f_{2}\!(i,\! j, \! k)$}%
    \put(0.53,0.45){$\! 1 \! - \! f_{1} \! ( \! i, \! j, \! k)$}%
     \put(0.785,0.1){${f_{2,p} \! (i \! + \! 1, \! j, \! k \! )}$}%
  \end{picture}%
\endgroup%

%% file: pictures/Advektion3fluid_V2.pdf_tex
\begingroup%
  \makeatletter%
  \providecommand\color[2][]{%
    \errmessage{(Inkscape) Color is used for the text in Inkscape, but the package 'color.sty' is not loaded}%
    \renewcommand\color[2][]{}%
  }%
  \providecommand\transparent[1]{%
    \errmessage{(Inkscape) Transparency is used (non-zero) for the text in Inkscape, but the package 'transparent.sty' is not loaded}%
    \renewcommand\transparent[1]{}%
  }%
  \providecommand\rotatebox[2]{#2}%
  \newcommand*\fsize{\dimexpr\f@size pt\relax}%
  \newcommand*\lineheight[1]{\fontsize{\fsize}{#1\fsize}\selectfont}%
  \ifx\svgwidth\undefined%
    \setlength{\unitlength}{99.21259843bp}%
    \ifx\svgscale\undefined%
      \relax%
    \else%
      \setlength{\unitlength}{\unitlength * \real{\svgscale}}%
    \fi%
  \else%
    \setlength{\unitlength}{\svgwidth}%
  \fi%
  \global\let\svgwidth\undefined%
  \global\let\svgscale\undefined%
  \makeatother%
  \begin{picture}(1,1)%
    \lineheight{1}%
    \setlength\tabcolsep{0pt}%
    \put(0,0){\includegraphics[width=\unitlength,page=1]{Advektion3fluid_V2.pdf}}%
    \put(0.39925327,0.80250332){\color[rgb]{1,0,0}\makebox(0,0)[lt]{\lineheight{1.25}\smash{\begin{tabular}[t]{l}\textbf{\textcolor{black}{$v \Delta t$}}\end{tabular}}}}%
    \put(0.75751516,0.66514893){\color[rgb]{0,0,0}\makebox(0,0)[lt]{\lineheight{1.25}\smash{\begin{tabular}[t]{l}\textbf{$flux_{1}$}\end{tabular}}}}%
    \put(0.02830767,0.66187832){\color[rgb]{0,0,0}\makebox(0,0)[lt]{\lineheight{1.25}\smash{\begin{tabular}[t]{l}\textbf{$flux_{2}$}\end{tabular}}}}%
  \end{picture}%
\endgroup%

%% file: 02_Validation.tex
\section{Validation} \label{Sec:Validation}%
The discussed modifications of the VOF method for immiscible liquids and their implementation into the software FS3D were successfully validated with analytical and experimental test cases as presented in the following. A liquid lens test case verifies the modeling of the surface forces, while comparisons of the morphology of binary droplet collisions of different liquid combinations from experiments validate the overall concept and its implementation. Quantitative comparisons of regime boundaries support this. Furthermore, the advection extended to three phases was verified within the framework of convergence studies. These studies are shown in \ref{App:AdvectTest} and \ref{App:ConvergenceEnergy}. The details on the computational infrastructure are given in \ref{App:ComputationalInfrastructure}.
\input{02_01_03_InterfacialTension}
%
\input{02_02_ComparisonExperiment}

%% file: 02_01_03_InterfacialTension.tex
\subsection{Liquid Lens as Three-phase Surface Force Test} \label{Subsec:SurfaceTest} %
\begin{figure*}[!htb]
\centering
\def\svgwidth{8cm}
\includegraphics[width=3.5cm]{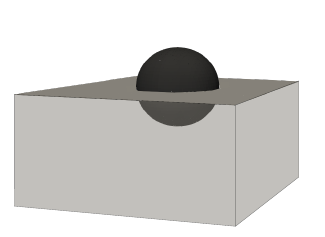}
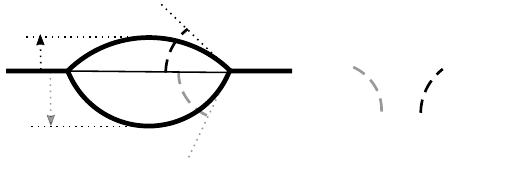
 \caption{Liquid lens test case: Initialization (left) of a spherical droplet of liquid~$1$ (dark grey) half dipped into a basin of liquid~$2$ (light grey) in a gaseous environment, $3$. %
After reaching steady-state, the lens shape is given by two intersecting spheres with base radius $a$ and spherical cap heights $h_{mn}$ (center) and defined by the relation of the surface- and interfacial tensions $\sigma_{mn}$ given by Neumann's triangle (right).} %
 \label{Fig:LiquidLensSketch}
\end{figure*}
\begin{figure*}[!tb]
\null\hfill
 \begin{minipage}[b]{0.33\linewidth}
 \flushleft
\includegraphics[scale=0.75,page=1,trim=0.2cm 0cm 0cm 0cm,clip]{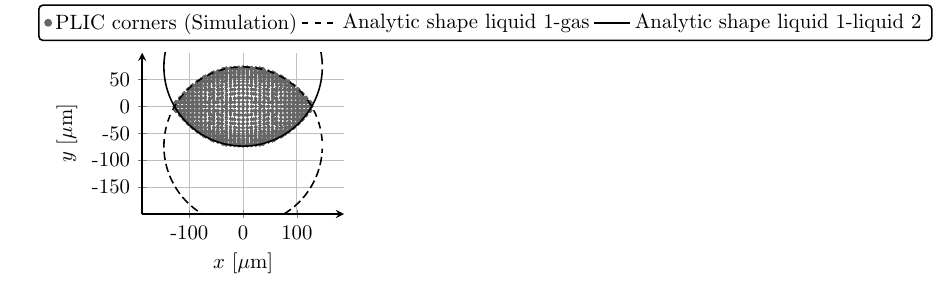}
\text{\small{$(\sigma_{13}, \sigma_{23},\sigma_{12})=(20,20,20)$}}
 \end{minipage}
 \hfill
 \begin{minipage}[b]{0.21\linewidth}
 \centering
\includegraphics[scale=0.75,page=2,trim=0.7cm 0cm 11cm 0cm,clip]{LiquidLensTests_New.pdf}
\text{\small{$(20,10,20)$}}
 \end{minipage}
 \hfill
 \begin{minipage}[b]{0.21\linewidth}
 \centering
\includegraphics[scale=0.75,page=3,trim=0.7cm 0cm 11cm 0cm,clip]{LiquidLensTests_New.pdf}
\text{\small{$(25,20,15)$}} 
 \end{minipage}
 \hfill
 \begin{minipage}[b]{0.21\linewidth}
 \centering
\includegraphics[scale=0.75,page=4,trim=0.7cm 0cm 11cm 0cm,clip]{LiquidLensTests_New.pdf}
\text{\small{\hspace{15pt}$(70,22,50) \mathrm{~\frac{mN}{m}}$}}
 \end{minipage}
 \hfill\null
\caption{Comparison of the rotational symmetric analytical solution in three dimensions for the interfaces and simulated stationary liquid lens shapes (grey, all three dimensional PLIC corner points projected onto the plane of view)  for an initial droplet radius of  liquid~1 of $R_{0,1}=100~\mathrm{~\mu m}$ resolved with $32~\mathrm{cells}/250~\mathrm{\mu m}$ in the simulations at different given surface and interfacial tensions.} \label{Fig:LensShapes}
\end{figure*}
A liquid lens is the resulting steady state (shown in a sectional view in Fig.~\ref{Fig:LiquidLensSketch} (center)) of a droplet of liquid~$1$ dipped half in a basin of liquid~$2$  surrounded by air, $3$, as shown in Fig.~\ref{Fig:LiquidLensSketch} (left). The shape of the liquid lens is determined by the interfacial tensions. It can be derived from geometrical considerations and the surface force balance given by Neumann's triangle, see Fig.~\ref{Fig:LiquidLensSketch} (right).
If the liquids' combination is fully wetting, Neumann's triangle is not closed any more and the liquid~$1$ droplet "drowns" in the liquid~$2$ bath due to the unbalanced surface forces. %
\ref{App:LiquidLens} provides the full derivation of the analytical solution of the three dimensional liquid lens's shape. %
The radius of the base circle reads
\begin{equation}
a =  R_{0,1}\left( \frac{4}{\phi_{12} + \phi_{13}} \right)^\frac{1}{3} 
\end{equation}
with the abbreviation
\begin{equation}
\phi_{mn} =  \frac{1}{\left(\sin(\theta_{mn})\right)^3} \left(2+\cos(\theta_{mn})\right)\left(1-\cos(\theta_{mn})\right)^2 
\end{equation}
and $R_{0,1}$ the radius of the initial sphere of liquid~$1$.
The height of each of the two spheroidal caps is
\begin{equation}
\begin{aligned}
h_{mn}   = a \tan{\frac{\theta_{mn}}{2}}\text{.}
\end{aligned}
\end{equation}
The angles $\theta_{mn}$ are depicted in Fig.~\ref{Fig:LiquidLensSketch}~(center and right).
Four simulation results with different interfacial- and surface tensions leading to  different lens shapes are shown in Fig.~\ref{Fig:LensShapes}. %
The diameter of the initially round droplet of liquid~1 was set to $200~\mu m$, the densities and viscosities of all phases were set equal to $1000~\frac{kg}{m^3}$ and $4~\mathrm{mPas}$. %
\begin{table}[!tb]
\caption{Relative errors of the lens dimensions for the simulated stationary state compared to the analytical lens dimensions for different combinations of interfacial tensions depicted in Fig.~\ref{Fig:LensShapes}. } \label{Tab:LensErrors}
\centering
 \begin{tabular}{l|lll}
 \hline
 \hline
 & & &  \\
\textbf{$\sigma_{13}$-$\sigma_{23}$-$\sigma_{12}$}  & $\Delta a_{\mathrm{rel}}$ & $\Delta h_{12,\mathrm{rel}}$ & $\Delta h_{13,\mathrm{rel}}$  \\
 $\mathrm{mN/m}$  & \%  & \% & \%  \\
 & & &  \\
 \hline
 & & &  \\
  $20$-$20$-$20$  &  $0.409$ & $0.020$ & $0.020$  \\
 $20$-$10$-$20$  &  $0.008$ & $0.111$ & $0.111$  \\
 $25$-$20$-$15$  &  $0.441$ & $0.136$ & $0.408$  \\ 
  $70$-$22$-$50$ &  $4.739$ & $0.141$ & $2.964$  \\
 & & &  \\
  \hline
  \hline
 \end{tabular}
\end{table}
The simulated lens shapes and the analytical solution agree very well in all cases, cf.\ Fig.~\ref{Fig:LensShapes}. %
Table~\ref{Tab:LensErrors} summarizes the errors of the height and width of the lens. %
The slightly higher errors  $\Delta h_{13,{\mathrm{rel}}}$ and $\Delta a_{\mathrm{rel}}$ for the close to fully wetting case $(\sigma_{13},\sigma_{23},\sigma_{12})=(70,22,50)$ are due to the low absolute height of the upper part of the liquid lens and the small lens's width. %
The absolute errors for this case are still all well below a cell's edge length. %

%% file: pictures/LiquidLensSketch_NeumannTriangle.pdf_tex
\begingroup%
  \makeatletter%
  \providecommand\color[2][]{%
    \errmessage{(Inkscape) Color is used for the text in Inkscape, but the package 'color.sty' is not loaded}%
    \renewcommand\color[2][]{}%
  }%
  \providecommand\transparent[1]{%
    \errmessage{(Inkscape) Transparency is used (non-zero) for the text in Inkscape, but the package 'transparent.sty' is not loaded}%
    \renewcommand\transparent[1]{}%
  }%
  \providecommand\rotatebox[2]{#2}%
  \newcommand*\fsize{\dimexpr\f@size pt\relax}%
  \newcommand*\lineheight[1]{\fontsize{\fsize}{#1\fsize}\selectfont}%
  \ifx\svgwidth\undefined%
    \setlength{\unitlength}{255.11811024bp}%
    \ifx\svgscale\undefined%
      \relax%
    \else%
      \setlength{\unitlength}{\unitlength * \real{\svgscale}}%
    \fi%
  \else%
    \setlength{\unitlength}{\svgwidth}%
  \fi%
  \global\let\svgwidth\undefined%
  \global\let\svgscale\undefined%
  \makeatother%
  \begin{picture}(1,0.33333333)%
    \lineheight{1}%
    \setlength\tabcolsep{0pt}%
    \put(0,0){\includegraphics[width=\unitlength,page=1]{LiquidLensSketch_NeumannTriangle.pdf}}%
    \put(0.71886325,0.07809912){\color[rgb]{0,0,0}\makebox(0,0)[lt]{\lineheight{1.25}\smash{\begin{tabular}[t]{l}$\sigma_{23}$\end{tabular}}}}%
    \put(0.82727364,0.23065491){\color[rgb]{0,0,0}\makebox(0,0)[lt]{\lineheight{1.25}\smash{\begin{tabular}[t]{l}$\sigma_{13}$\end{tabular}}}}%
    \put(0.5931374,0.21804491){\color[rgb]{0,0,0}\makebox(0,0)[lt]{\lineheight{1.25}\smash{\begin{tabular}[t]{l}$\sigma_{12}$\end{tabular}}}}%
    \put(0.65273488,0.13765625){\color[rgb]{0,0,0}\makebox(0,0)[lt]{\lineheight{1.25}\smash{\begin{tabular}[t]{l}$\theta_{12}$\end{tabular}}}}%
    \put(0.82366231,0.13065073){\color[rgb]{0,0,0}\makebox(0,0)[lt]{\lineheight{1.25}\smash{\begin{tabular}[t]{l}$\theta_{13}$\end{tabular}}}}%
    \put(0.34796744,0.16270421){\color[rgb]{0,0,0}\makebox(0,0)[lt]{\lineheight{1.25}\smash{\begin{tabular}[t]{l}$\theta_{12}$\end{tabular}}}}%
    \put(0.33335578,0.2135696){\color[rgb]{0,0,0}\makebox(0,0)[lt]{\lineheight{1.25}\smash{\begin{tabular}[t]{l}$\theta_{13}$\end{tabular}}}}%
    \put(0.03974724,0.13633332){\color[rgb]{0,0,0}\makebox(0,0)[lt]{\lineheight{1.25}\smash{\begin{tabular}[t]{l}$h_{12}$\end{tabular}}}}%
    \put(0.01166979,0.21944922){\color[rgb]{0,0,0}\makebox(0,0)[lt]{\lineheight{1.25}\smash{\begin{tabular}[t]{l}$h_{13}$\end{tabular}}}}%
    \put(0,0){\includegraphics[width=\unitlength,page=2]{LiquidLensSketch_NeumannTriangle.pdf}}%
    \put(0.26717781,0.03389085){\color[rgb]{0,0,0}\makebox(0,0)[lt]{\lineheight{1.25}\smash{\begin{tabular}[t]{l}$2 a$\end{tabular}}}}%
    \put(0,0){\includegraphics[width=\unitlength,page=3]{LiquidLensSketch_NeumannTriangle.pdf}}%
  \end{picture}%
\endgroup%

%% file: 02_02_ComparisonExperiment.tex
\subsection{Comparison with Experiments} \label{Subsec:ExperimentCompare} %
 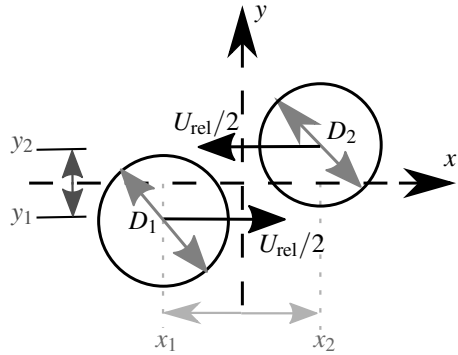
\begin{figure}[!htb]
 \centering
\setlength{\unitlength}{6cm}
\begin{picture}(1,0.9)
\put(0,0){\includegraphics[width=\unitlength]{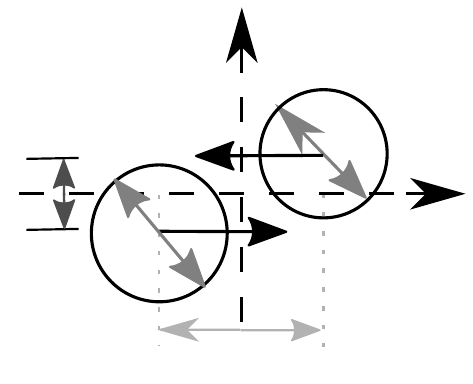}}
    \put(0.55,0.25){$U_\mathrm{rel}/2$}%
    \put(0.36,0.52){$U_\mathrm{rel}/2$}%
    \put(0.26,0.3){$D_1$}%
    \put(0.7,0.5){$D_2$}%
    \put(0.32,0.04){\textcolor{black!60!}{$x_1$}}%
    \put(0.67,0.04){\textcolor{black!60!}{$x_2$}}%
    \put(0.0,0.32){\textcolor{black!70!}{$y_1$}}%
    \put(0.0,0.48){\textcolor{black!70!}{$y_2$}}%
    \put(0.96,0.44){$x$}%
    \put(0.54,0.77){$y$}%
\end{picture}
    \caption{Simulation setup in the $x-y$-plane for the droplet collisions. The perpendicular $z$-axis is not shown. %
    At $z=0$, a symmetry plane is employed to reduce the domain size for the simulations. %
    If exactly head-on collisions are computed ($y_m=0$), an additional symmetry plane at $y=0$ can be introduced.} \label{Fig:SetupCollision}
 \end{figure}
 \begin{figure}
 \centering
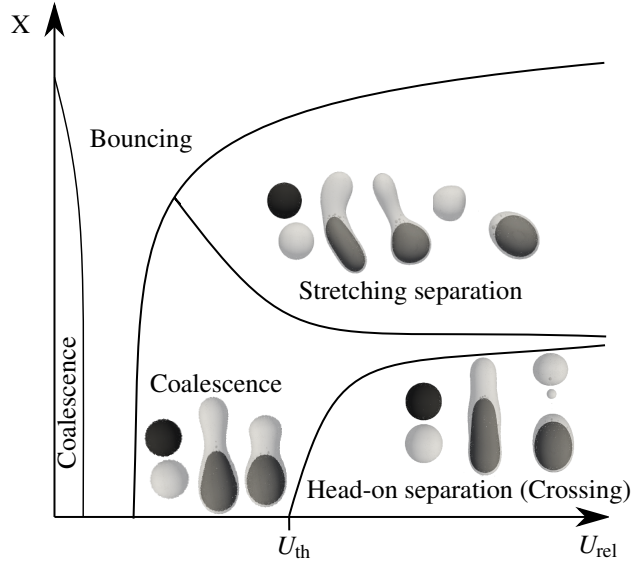
\caption{Exemplary regime map characterizing the outcome of binary collisions of immiscible droplets. This study focuses on the regimes close to the head-on separation threshold illustrated with examples.} %
\label{Fig:ExampleRegimemap}
 \end{figure}
A comparison with experiments on immiscible droplet collisions by \citet{Planchette2011} reveals a comprehensive picture of the quality of the simulation results. Depending on the relative offset of the droplet centers, the impact parameter,
\begin{equation}
X= \frac{y_1+y_2}{0.5 \left(D_1+D_2\right)}\text{,}
\end{equation}
and the relative collision velocity, $U_\mathrm{rel}$, cf.\ Fig.~\ref{Fig:SetupCollision}, different collision outcomes are possible: Bouncing, head-on as well as off-center coalescence or separation. An exemplary regime map is sketched in Fig.~\ref{Fig:ExampleRegimemap}. In accordance to the experimental data, we compare to in the following, the dimensional velocity was chosen as horizontal axis. %
At small impact parameters and low velocities beyond the bouncing regime that is not considered in the present study, the droplets will coalesce, see Figs.~\ref{Fig:G50SOM5-X0023Urel317} and \ref{Fig:G50SOM5-X021Urel330}. If the impact parameter increases, stretching separation takes place independent of the chosen liquids' combination, see Fig.~\ref{Fig:G50SOM5-X033Urel298}. %
At higher velocities and low impact parameters, close to head-on collisions, the separation mechanism depends on the chosen liquids: %
Crossing \citep{Planchette2009, Planchette2011}, single reflex \citep{Chen2006a,Planchette2009,Planchette2011} or reflexive separation \citep{Planchette2011, Planchette2012} are the possible outcomes. Crossing separation results in one compound droplet and one pure droplet of liquid~2 on the rear side (Fig.~\ref{Fig:G50SOM5-X0057Urel346} and \ref{Fig:G50SOM5-X0075Urel330}), while after reflexive separation the pure liquid~2 droplet is reflected back (Fig.~\ref{Fig:G50Perfluo}). The result of single reflex separation are two compound droplets (Fig.~\ref{Fig:G50SOM510}).  The combinations of liquids used in the experimental study we compare to are always fully wetting, which leads to a full encapsulation of the inner liquid in all cases.\\
Since there is little published experimental data on the collision of immiscible liquids  beyond the work of \citet{Planchette2011}, as outlined in Sec.~\ref{Sec:Introduction}, we use this data to validate our simulations. We were kindly provided with further unpublished pictures as well as precise details of the setup for this purpose by C. Planchette. 
This significantly improves the reliability of the comparisons, as a first feasibility study revealed  that even small deviations in the fluid properties or geometry data in the simulation setup have a significant influence on the regime boundary. \\
Table~\ref{Tab:Properties} summarizes the liquids' properties for the investigated collisions. A 50$\%$-glycerol-water solution (G50) is used as liquid 1, liquid 2 is a silicon oil with varying viscosity (SOMx) or Perfluodecaline (Perfluo).
\begin{table}[!tb]
 \caption{Properties of liquid combinations for the collisions of G50 (index $1$) and SOM5, SOM5+10, SOM3 or Perfluo (index $2$) droplets  given by \citet{Planchette2011} and properties of air (index $3$) utilized in the simulations. %
} \label{Tab:Properties}%
\centering
\begin{tabular}{lllll}
\hline
\hline
& & & &  \\
Liquid/gas ($m$)  & $\mu_m$ & $\rho_m$  &  $\sigma_{m3}$ & $\sigma_{12}$ \\
& $\mathrm{mPas}$ & $\mathrm{\frac{kg}{m^3}}$ & $\mathrm{\frac{mN}{m}}$ & $\mathrm{\frac{mN}{m}}$ \\
                                 &          &              &         &                               \\
 \hline
                                 &          &              &         &                               \\
G50 ($1$)   & 6       &  1126  & 68.6 & cf.\ ($2$)  \\
                                 &          &              &         &                               \\
SOM3 ($2$)         & 2.79 &   892.2  & 19.5 & 34.9   \\                                 
SOM5 ($2$)         & 4.57 &  913.4  & 19.5 & 34.3  \\
SOM5+10 ($2$)  & 6.6   & 925.3  & 19.8 & 34.5   \\
Perfluo ($2$)       & 5.5   & 1934.9 & 17.8 & 36.5  \\
                                 &          &              &         &                               \\
air ($3$)                  &  0.01824 & 1.19 & cf.\ ($1$)\&($2$) & -   \\                                 
                                 &          &              &         &                               \\
 \hline
 \hline
 \end{tabular}
\end{table}

The simulation is set up according to Fig.~\ref{Fig:SetupCollision}. In case of head-on collisions, $X=0$, two symmetry conditions at $y=0$ and $z=0$ were applied in the simulation in order to quarter the size of the computational domain. Off-center collisions were performed with one symmetry plane at $z=0$. The other boundaries were set to slip walls, which allows the evaluation of the energy budget in a closed system. Evaluations close to the head-on boundary reveal, that the domain size has no noticeable influence on the collisions' outcome as long as the collision complex does not touch the domain's boundary at maximum extension. %
The domain was adjusted depending on the expected outcome in order to reach a slightly higher resolution, if possible, to better resolve thin ligaments and films as well as to save computational resources. Thus, the resolution ranges from from $768~\mathrm{cells}/\mathrm{mm}$ to $960~\mathrm{cells}/\mathrm{mm}$. The convergence study for the head-on boundary as well as the energy budget shown in \ref{App:Convergence} reveals that the chosen resolutions are adequate for the simulated cases.

%
\subsubsection{Comparison of Droplet Collisions of 50\%-Glycerol Solution (G50) and Silicon Oil M5 (SOM5)}  \label{Subsec:G50SOM5Coll}
%
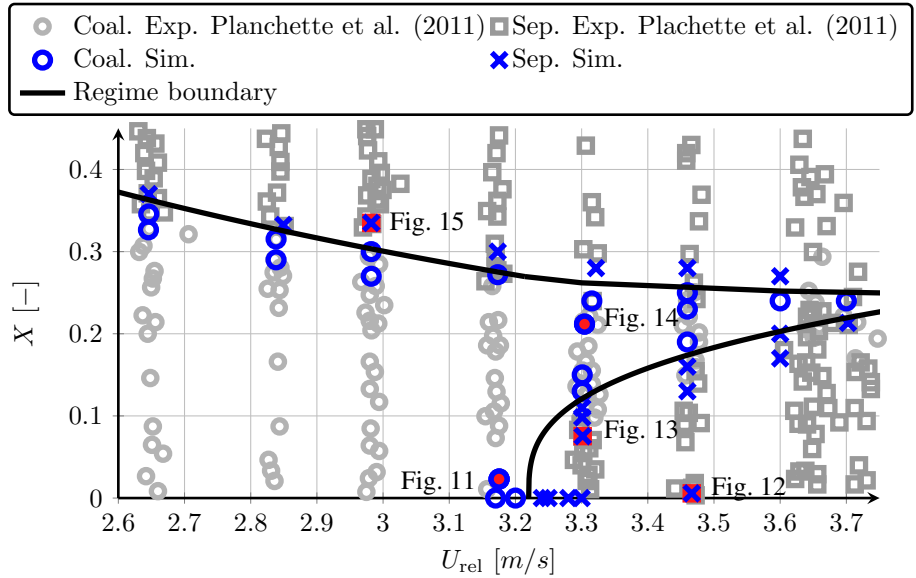
\begin{figure*}[!htbp]
	\centering
	\setlength{\unitlength}{15cm}
	\begin{picture}(1,0.57)
	\put(0,0){\includegraphics[scale=1]{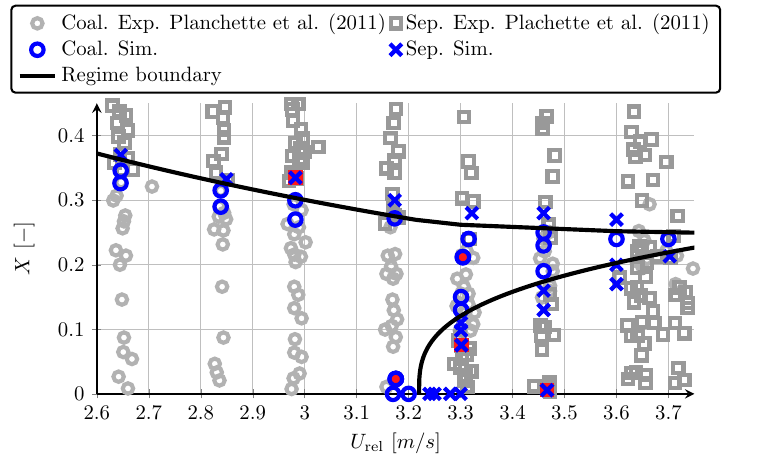}}
	\put(0.36,0.09){Fig.~\ref{Fig:G50SOM5-X0023Urel317}}
	\put(0.635,0.085){Fig.~\ref{Fig:G50SOM5-X0057Urel346}}
	\put(0.54,0.135){Fig.~\ref{Fig:G50SOM5-X0075Urel330}}
	\put(0.54,0.235){Fig.~\ref{Fig:G50SOM5-X021Urel330}}
	\put(0.35,0.32){Fig.~\ref{Fig:G50SOM5-X033Urel298}}
	\end{picture}
	\caption{Regime map of collisions of G50-SOM5: %
		Comparison of our simulations with experiments by \citet{Planchette2011}. %
		For the points marked in red, the comparison of the morphology is shown in \mbox{Figs.~\ref{Fig:G50SOM5-X0023Urel317} --~\ref{Fig:G50SOM5-X033Urel298}}. %
	}\label{Fig:RegimeMapSOM5G50}
\end{figure*}
\begin{figure*}[!htbp]
\centering
 \includegraphics[width=\textwidth, trim=0cm 0cm 0cm 0.4cm,clip]{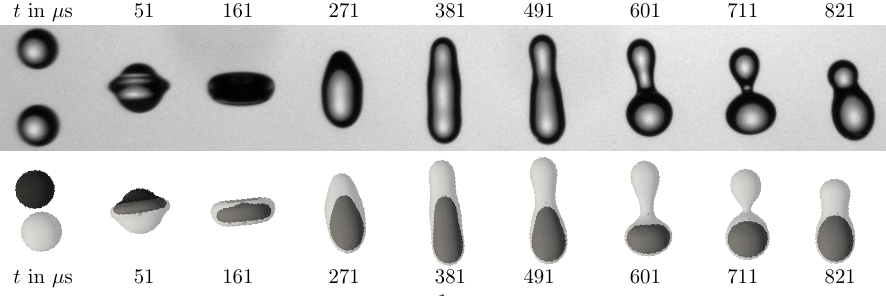}
 \caption{Head-on coalescence: Comparison of shadow images from experiments (top) and the simulation (bottom) of the collision of G50 (dark grey) and SOM5 (transparent light grey): $X=0.023$, $U_\mathrm{rel}=3.17~\mathrm{m/s}$, $D_1= 189 ~\mathrm{\mu m}$, $D_2=196 ~\mathrm{\mu m}$, $\Delta t=110~\mathrm{\mu s}$, $768~\mathrm{cells}/0.8~\mathrm{mm}$. %
The exact setup and picture of experiment was provided by courtesy of C. Planchette. } \label{Fig:G50SOM5-X0023Urel317}
\end{figure*}
\begin{figure*}[!htbp]
\centering
 \includegraphics[width=\textwidth, trim=0cm 0cm 0cm 0.35cm,clip]{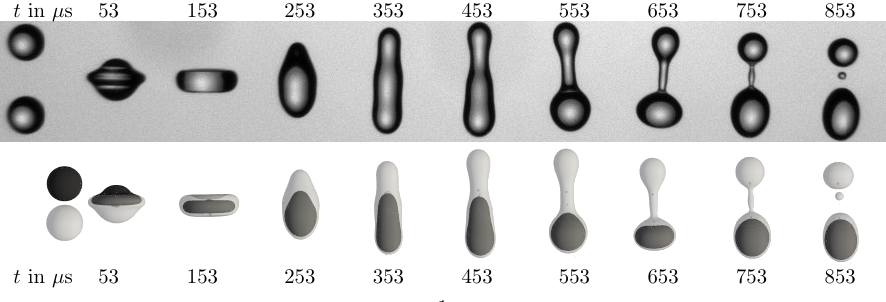}
 \caption{Head-on crossing separation: Comparison of shadow images from experiments (top) and the simulation (bottom) of the collision of G50 (dark grey) and SOM5 (transparent light grey): $X=0.057$, $U_\mathrm{rel}=3.46~\mathrm{m/s}$, $D_1= 188~\mathrm{\mu m}$, $D_2= 195~\mathrm{\mu m}$, $\Delta t=100~\mathrm{\mu s}$,  $768~\mathrm{cells}/1~\mathrm{mm}$. %
 The exact setup and picture of experiment was provided by courtesy of C. Planchette.}\label{Fig:G50SOM5-X0057Urel346}
\end{figure*}
\begin{figure*}[!htbp]
\centering
 \includegraphics[width=\textwidth, trim=0cm 0cm 0cm 0.38cm,clip]{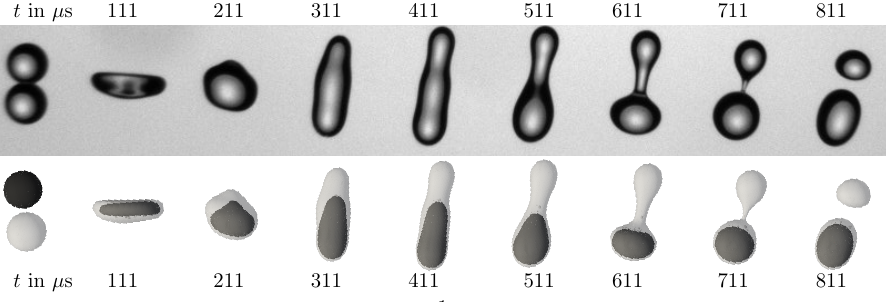}
 \caption{Close to head-on crossing separation: Comparison of shadow images from experiments (top) and the simulation (bottom) of the collision of G50 (dark grey) and SOM5 (transparent light grey): $X=0.075$, $U_\mathrm{rel}=3.30~\mathrm{m/s}$, $D_1= 189~\mathrm{\mu m}$, $D_2= 195~\mathrm{\mu m}$, $\Delta t=100~\mathrm{\mu s}$,  $768~\mathrm{cells}/0.8~\mathrm{mm}$. The exact setup and picture of experiment was provided by courtesy of C. Planchette.}\label{Fig:G50SOM5-X0075Urel330}
\end{figure*}
\begin{figure*}[!htbp]
\centering
 \includegraphics[width=\textwidth, trim=0cm 0cm 0cm 0.4cm,clip]{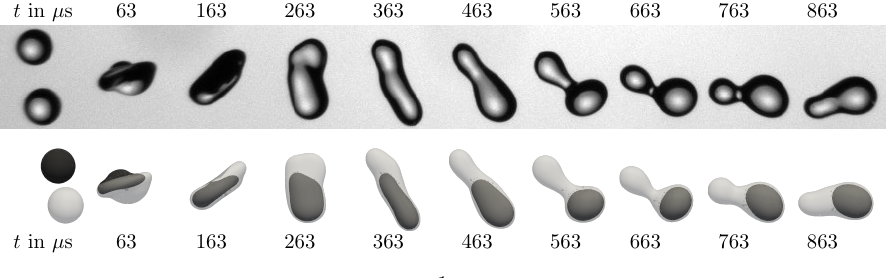}
 \caption{Off-center coalescence: Comparison of shadow images from experiments (top) and the simulation (bottom) of the collision of G50 (dark grey) and SOM5 (transparent light grey): %
 $X=0.21$, $U_\mathrm{rel}=3.3~\mathrm{m/s}$, $D_1=189~\mathrm{\mu m}$, $D_2=196~\mathrm{\mu m}$, $\Delta t=100~\mathrm{\mu s}$, $768~\mathrm{cells}/0.8~\mathrm{mm}$. %
 The exact setup and picture of experiment was provided by courtesy of C. Planchette.}\label{Fig:G50SOM5-X021Urel330}
\end{figure*}
\begin{figure*}[!htbp]
\centering
 \includegraphics[width=\textwidth, trim=0cm 0cm 0cm 0.4cm,clip]{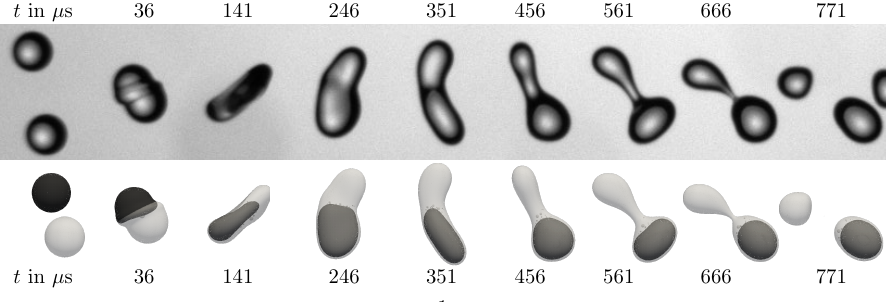}
 \caption{Stretching separation: Comparison of shadow images from experiments (top) and the simulation (bottom) of the collision of G50 (dark grey) and SOM5 (transparent light grey): %
 $X=0.33$, $U_\mathrm{rel}=2.98~\mathrm{m/s}$, $D_1=189~\mathrm{\mu m}$, $D_2=195~\mathrm{\mu m}$, $\Delta t = 105~\mathrm{\mu s}$, $768~\mathrm{cells}/0.8~\mathrm{mm}$. %
 The exact setup and picture of experiment was provided by courtesy of C. Planchette.}\label{Fig:G50SOM5-X033Urel298}
\end{figure*}
The interaction of SOM5 and G50 was chosen for a comparison of the coalescence and separation regimes. The resulting regime map reveals excellent agreement between the simulation and the experimental data points, with an accuracy below the experimental scatter, cf. Fig.~\ref{Fig:RegimeMapSOM5G50}. The simulations were initialized according to the information available from the experiments. Simulation points in the regime map with no exact geometrical information available were initialized with droplet diameters from reasonable close data points in the experimental dataset provided by C. Planchette to supplement the information published in \citet{Planchette2011}. %
A full list of the simulation setups and collision outcomes is available at \href{https://doi.org/10.18419/darus-3557}{https://doi.org/10.18419/darus-3557}. %

The morphology of the collisions was compared for selected cases that are marked in red in Fig.~\ref{Fig:RegimeMapSOM5G50}. Figures~\ref{Fig:G50SOM5-X0023Urel317}--\ref{Fig:G50SOM5-X0075Urel330} represent cases close to head-on collisions: Coalescence is shown in Fig.~\ref{Fig:G50SOM5-X0023Urel317} and crossing separation in Fig.~\ref{Fig:G50SOM5-X0057Urel346} and Fig.~\ref{Fig:G50SOM5-X0075Urel330}. %
Figure~\ref{Fig:G50SOM5-X021Urel330} shows a coalescence case at a larger offset, while a simulation of off-center stretching separation at a relatively low relative velocity is compared to the experimental results in Fig.~\ref{Fig:G50SOM5-X033Urel298}. Cases representing different collision outcomes are evaluated more detailed in Sec.~\ref{Sec:Results}. 
The comparisons in Figs.~\ref{Fig:G50SOM5-X0023Urel317}--\ref{Fig:G50SOM5-X033Urel298} each show excellent agreement between the experimental result in the top row and the simulation depicted below, not only in morphology, but also on the time scale. 
The G50-SOM5 experiments were performed with a droplet release frequency of $10 \pm 1$~kHz, the time step between two images is therefore $\Delta t \approx 91-111 ~\mu s$. The precise times in Figs.~\ref{Fig:G50SOM5-X0023Urel317}--\ref{Fig:G50SOM5-X033Urel298} were determined from the simulations. An equidistant output time step was chosen according to the specified frequency range at which there was the best morphological agreement with the first and last image in the experiment.

\subsubsection{Comparison to Head-on Collisions of Other Liquid Combinations} \label{Subsec:head-onComparision}
\begin{figure*}[!htb]
\centering
\includegraphics[width=\textwidth, trim=0cm 0cm 0cm 0.55cm,clip]{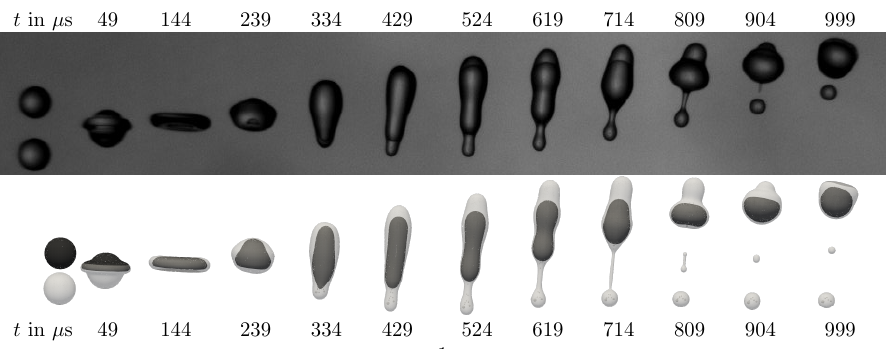}
\caption{Reflexive separation: Comparison of shadow images from experiments (top) and the simulation (bottom) of the collision of
	G50 (dark grey) and Perfluo (transparent light grey): $X=0.02$, $U_\mathrm{rel}=3.41~\mathrm{\frac{m}{s}}$, $D_1=183 \mathrm{\mu m}$, $D_2=185 \mathrm{\mu m}$, $\Delta t= 95~\mathrm{\mu s}$,  $1024~\mathrm{cells}/1.2~\mathrm{mm}$. The exact setup and picture of experiment was provided by courtesy of C. Planchette.} \label{Fig:G50Perfluo}
\end{figure*}

\begin{figure*}
\centering
\includegraphics[width=\textwidth,trim=0cm 0.5cm 0cm 0cm]{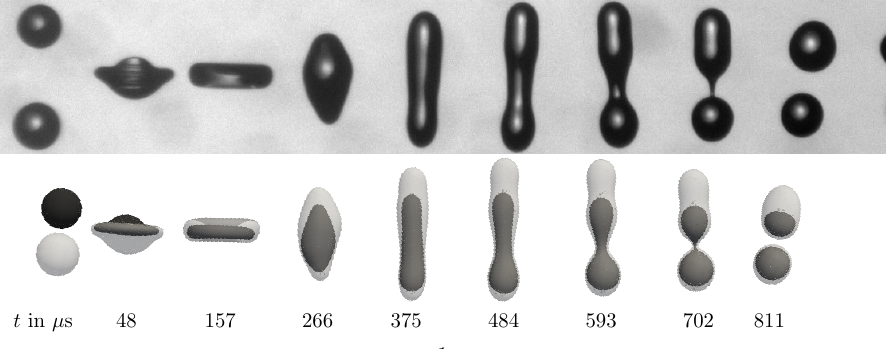}
\caption{Single reflex separation: Comparison of shadow images from experiments (top) and the simulation (bottom) of the collision of G50 (dark grey) and SOM5+10 (transparent light grey):  $X=0.06$, $U_\mathrm{rel}=4.52~\mathrm{m/s}$, $D_1=186~\mathrm{\mu m}$, $D_2=199~\mathrm{\mu m}$, $\Delta t = 109~\mathrm{\mu s}$, $768~\frac{\mathrm{cells}}{\mathrm{mm}}$. %
The exact setup and picture of experiment was provided by courtesy of C. Planchette. }
\label{Fig:G50SOM510}
\end{figure*}

In addition to the G50-SOM5 collisions, simulations of binary head-on droplet collisions of other liquid combinations, namely G50-SOM3 with a lower viscosity, G50-SOM5+10 (a mixture of SOM5 and SOM10) with a higher viscosity and G50-Perfluo with a higher density, were compared to experimental results. The head-on separation mechanisms crossing separation for G50-SOM3 and G50-SOM5, single reflex separation for G50-SOM5+10 and reflexive separation for G50-Perfluo are predicted by the simulation in accordance with the experiments. Also the corresponding threshold velocities, $U_{th}$, that indicate the transition from coalescence to separation and are iteratively found with the simulations are well within the experimental bounds by \citet{Planchette2009} and \citet{Planchette2011,Planchette2012,Planchette2017} as shown in Tab.~\ref{Tab:SummaryHeadOn}. Again, the morphology and time scales of the collision agrees well as shown for the reflexive separation case in Fig.~\ref{Fig:G50Perfluo} of G50-Perfluo and the single reflex separation case of the G50-SOM5+10 combination in Fig.~\ref{Fig:G50SOM510}. %
Both cases are close to the regime boundary of coalescence and separation for the respective liquids' combination. %
The largest deviations of the morphology are present for the G50-Perfluo collision depicted in Fig.~\ref{Fig:G50Perfluo}.
After around $t=500~\mu s$ the ligament of Perfluo is stretched longer than in the experiment, while the shape of the larger compound droplet agrees well with the experimental data. %
The reason lies in a still locally low resolution of the ligament, which is resolved with fewer cells than required for the stencil to compute the surface forces. %
Multiple enclosed small air bubbles add to this problem of premature ligament breakdown. %
A convergence study revealed, that the ligament becomes shorter with increased resolution, but fully resolving the ligament would result in infeasible compute times due to the reduced time step -- without significant further insights. Since the collision mechanism, the threshold velocity as well as the size and shape of the resulting droplets still agree well, no further refinement was employed. %


\begin{table}
\caption{Comparison of threshold velocities, $U_{th}$, separating the coalescence and separation head-on regime from simulation (sim) and experiment (exp) by \citet{Planchette2009} and \citet{Planchette2011,Planchette2012,Planchette2017}. %
The intervals are limited by the last velocity resulting in coalescence and the first resulting in separation.}\label{Tab:SummaryHeadOn}
\centering
 \begin{tabular}{lllll}
 \hline
 \\
  Liquid~$1$ & Liquid~$2$ & $U_\mathrm{th,exp}$ & $U_\mathrm{th,sim}$ & Separation  \\
            &             & $\mathrm{[m/s]}$     & $\mathrm{[m/s]}$     & mechanism \\
            \\
  \hline
  \\
  G50 & SOM3         &   2.56-2.57     & 2.50-2.57  & Crossing \\
  G50 & SOM5         &   3.18-3.30     & 3.20-3.24  & Crossing \\
  G50 & SOM5+10  &   3.81-4.70     &  3.90-4.52  & Single Reflex \\
  G50 & Perfluo      &   3.21-3.41     &  3.25-3.27  & Reflexive  \\
  \\
  \hline
  \hline
 \end{tabular} 
\end{table}

%% file: pictures/ExampleRegimeMap.pdf_tex
\begingroup%
  \makeatletter%
  \providecommand\color[2][]{%
    \errmessage{(Inkscape) Color is used for the text in Inkscape, but the package 'color.sty' is not loaded}%
    \renewcommand\color[2][]{}%
  }%
  \providecommand\transparent[1]{%
    \errmessage{(Inkscape) Transparency is used (non-zero) for the text in Inkscape, but the package 'transparent.sty' is not loaded}%
    \renewcommand\transparent[1]{}%
  }%
  \providecommand\rotatebox[2]{#2}%
  \newcommand*\fsize{\dimexpr\f@size pt\relax}%
  \newcommand*\lineheight[1]{\fontsize{\fsize}{#1\fsize}\selectfont}%
  \ifx\svgwidth\undefined%
    \setlength{\unitlength}{226.77165354bp}%
    \ifx\svgscale\undefined%
      \relax%
    \else%
      \setlength{\unitlength}{\unitlength * \real{\svgscale}}%
    \fi%
  \else%
    \setlength{\unitlength}{\svgwidth}%
  \fi%
  \global\let\svgwidth\undefined%
  \global\let\svgscale\undefined%
  \makeatother%
  \begin{picture}(1,0.95)%
    \lineheight{1}%
    \setlength\tabcolsep{0pt}%
    \put(0,0){\includegraphics[width=\unitlength,page=1]{ExampleRegimeMap.pdf}}%
    \put(0.23752391,0.29361162){\color[rgb]{0,0,0}\makebox(0,0)[lt]{\lineheight{1.25}\smash{\begin{tabular}[t]{l}Coalescence\end{tabular}}}}%
    \put(0.13755544,0.70006056){\color[rgb]{0,0,0}\makebox(0,0)[lt]{\lineheight{1.25}\smash{\begin{tabular}[t]{l}Bouncing\end{tabular}}}}%
    \put(0.11343746,0.16556416){\color[rgb]{0,0,0}\rotatebox{90}{\makebox(0,0)[lt]{\lineheight{1.25}\smash{\begin{tabular}[t]{l}Coalescence\end{tabular}}}}}%
    \put(0.48458061,0.44801345){\color[rgb]{0,0,0}\makebox(0,0)[lt]{\lineheight{1.25}\smash{\begin{tabular}[t]{l}Stretching separation\end{tabular}}}}%
    \put(0,0){\includegraphics[width=\unitlength,page=2]{ExampleRegimeMap.pdf}}%
    \put(0.49817829,0.11700125){\color[rgb]{0,0,0}\makebox(0,0)[lt]{\lineheight{1.25}\smash{\begin{tabular}[t]{l}Head-on separation (Crossing)\end{tabular}}}}%
    \put(0.00651133,0.89014581){\color[rgb]{0,0,0}\makebox(0,0)[lt]{\lineheight{1.25}\smash{\begin{tabular}[t]{l}X\end{tabular}}}}%
    \put(0.45,0.02347925){\color[rgb]{0,0,0}\makebox(0,0)[lt]{\lineheight{1.25}\smash{\begin{tabular}[t]{l}$U_\mathrm{th}$\end{tabular}}}}%
    \put(0,0){\includegraphics[width=\unitlength,page=3]{ExampleRegimeMap.pdf}}%
    \put(0.95096283,0.02501793){\color[rgb]{0,0,0}\makebox(0,0)[lt]{\lineheight{1.25}\smash{\begin{tabular}[t]{l}$U_\mathrm{rel}$\end{tabular}}}}%
  \end{picture}%
\endgroup%

%% file: 03_Results.tex
\section{Results and Discussion} \label{Sec:Results}%
Analytical models of collision outcomes are usually based on geometrical considerations and an evaluation of energy budgets, cf. Sec.~\ref{Sec:Introduction}, which is not accessible in such detail in the experiments. The kinetic, surface and dissipated energy are evaluated along with the viscous dissipation rate for each time-step in our simulations of selected collisions at run-time in the following. These are the cases marked in red in Fig.~\ref{Fig:RegimeMapSOM5G50}. %
\ref{App:Energy} presents the details on the computation of the evaluated terms. The evaluation not only shows that the implemented method is energy conserving, with losses below 4$\%$ for the employed resolutions, see \ref{App:Convergence}, it also provides insights that are substantial for future modeling. A comparison to single liquid collisions supports this by revealing the corresponding similarities and differences.

\subsection{Energy Budgets of Selected Droplet Collision Cases} \label{Subsec:EnergyBalance}
\begin{figure*}
\centering
\includegraphics[scale=0.75,trim=0cm 0.4cm 0cm 0cm,clip]{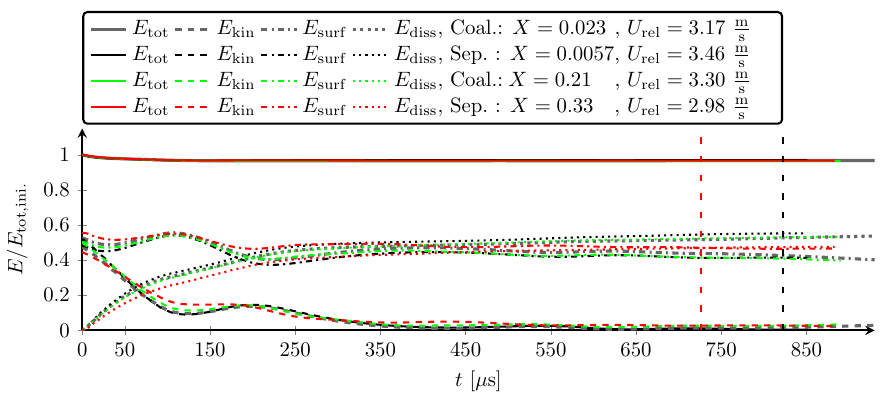}
\caption{Energy budgets of representative G50-SOM5 droplet collisions, which result in coalescence or separation. %
The time of separation is marked by a vertical dashed line in the respective color. Stretching separation appears earlier than head-on crossing separation despite the lower velocity. %
} \label{Fig:G50-SOM5_EnergyBudgets}
\end{figure*}

\begin{figure*}
	\centering
	\includegraphics[scale=0.75,page=1]{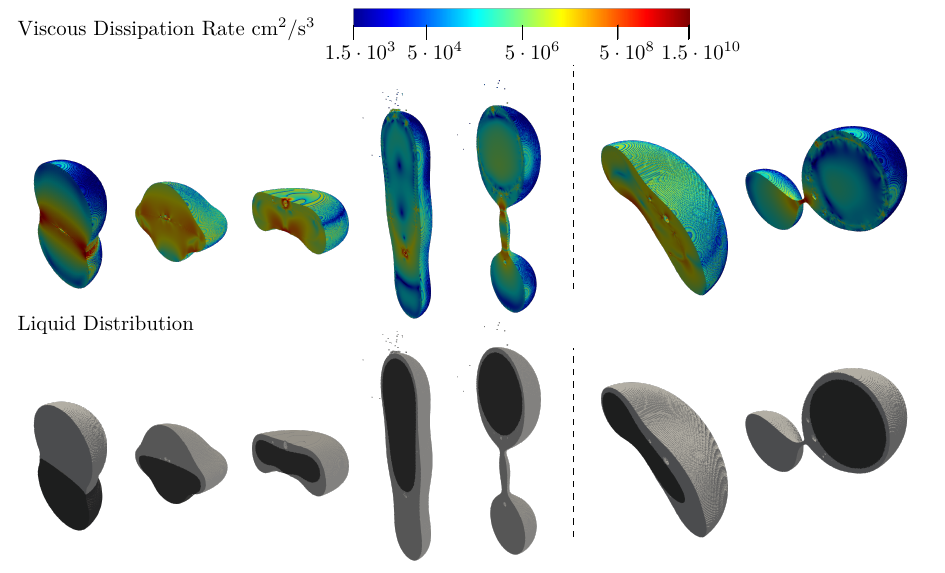}
	\caption{Distribution of the local viscous dissipation rate, $\Phi_\mathrm{cell}$, cf. \ref{subsec:DissipatedEnergy} from a head-on separation (five pictures left) and an off-center separation (two pictures right) is shown with hotspots at thin films, triple lines and at thin ligaments. The phase distribution liquid~1 (dark grey) and liquid~2 (light grey) is illustrated below.} \label{Fig:DissipationHotspots}
\end{figure*}

Already during the approach of the droplets, deformations in the surface of the outer liquid and an energy transfer are present, which becomes visible by, e.g., a gradient in the surface energy. After contact, the surface energy increases towards a maximum, which appears slightly earlier than the minimum in kinetic energy. This is true for all investigated collisions. In general, a comparison of the energy budgets for head-on and off-center coalescence as well as stretching and crossing separation reveals no major differences or characteristic features. A minor difference becomes visible in the case of stretching separation with a high impact parameter of \mbox{$X~=~0.33$}, which, in contrast to the other cases, does not reveal an expressed minimum in the normalized kinetic energy, but a plateau. Furthermore, the second surface energy minimum is on a higher level. Another interesting finding is that, despite the higher initial relative velocity, the separation time is later for the crossing separation case than the stretching separation case, which is indicated by a vertical dashed line in the respective color in Fig.~\ref{Fig:G50-SOM5_EnergyBudgets}.\\
As shown in Fig.~\ref{Fig:DissipationHotspots}, the Viscous Dissipation Rate (VDR) has a maximum at the initial stage of the collision, which originates from a maximum in the velocity and viscosity gradients along the thin gas film between the approaching droplets. %
It also takes on high values on the rear side of the collision complex, see Fig.~\ref{Fig:DissipationHotspots} (third left), where the triple line of the encapsulating liquid meets from all sides and forms a new bulk of liquid~$2$, cf. Fig.~\ref{Fig:DissipationHotspots} (fourth left). %
This effect is prominent in head-on cases as the inertia is distributed equally around the encapsulated droplet's perimeter. %
In case of stretching separation on the other hand, the initial offset of the droplets leads to most of the mass moving along one side of the inner droplet. %
The local velocity gradients and the local VDR are thus lower at the rear side of the collision, cf. Fig.~\ref{Fig:DissipationHotspots} (second right). %
The later stages of the collision show, that high VDRs are not only present at the triple line, but also at thin ligaments shown in Fig.~\ref{Fig:DissipationHotspots}, both for head-on crossing (third from the right) and stretching separation (right).\\

 \subsection{Comparison with Single Liquid Collisions} \label{Subsec:SingleLiquidComparison}
\begin{figure}[!htb]
  \centering
   \includegraphics[scale=0.75]{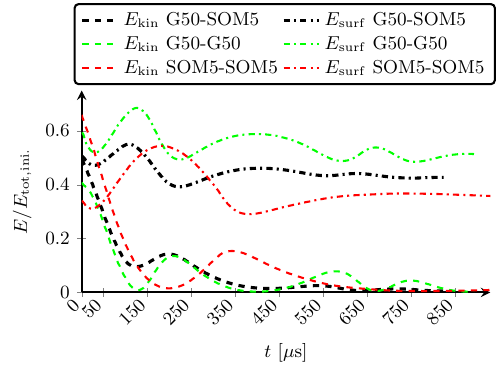}
	\caption{A comparison of the collision of G50-SOM5 resulting in crossing separation with collisions of the single liquids resulting in coalescence ($X~=~0.075$ and $U_\mathrm{rel}~=~3.3~\mathrm{m/s}$) indicates that the behaviour of the inner phase initially determines the collision.}\label{Fig:OsciallationCompare}
\end{figure}

\begin{figure*}
   \centering
    \includegraphics[scale=0.75]{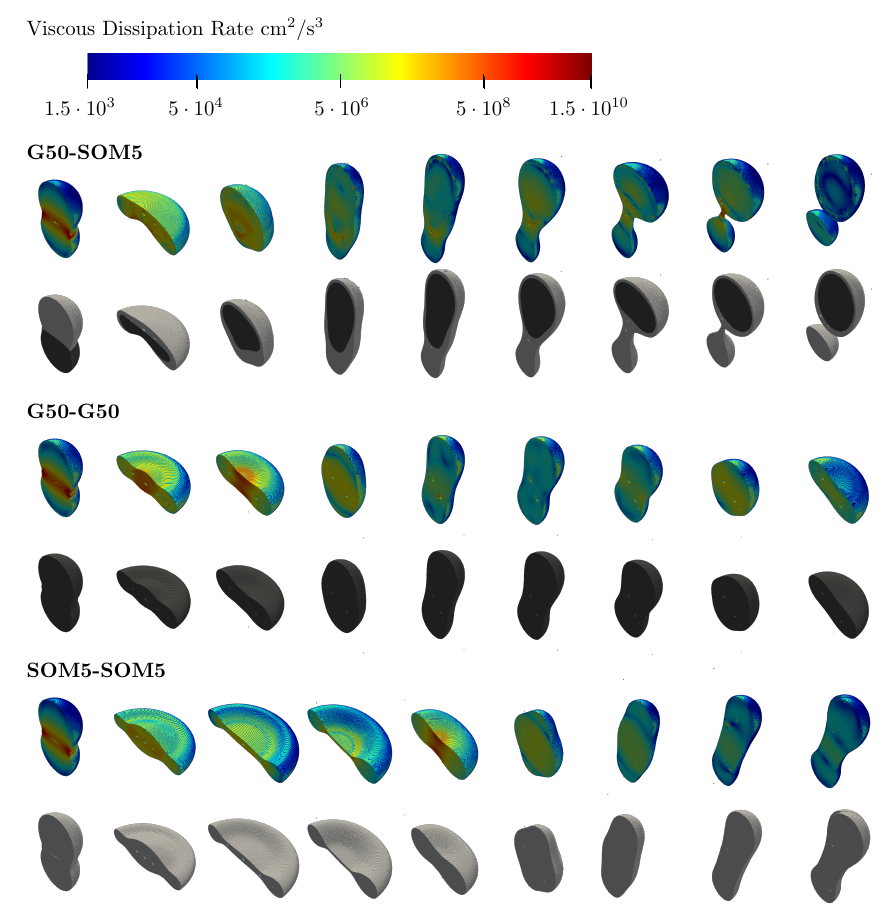}  
  \caption{Comparison of crossing separation of SOM5-G50 and the respective single liquid collisions by the liquid distribution and viscous dissipation rate for the same time steps like in Fig.~\ref{Fig:G50SOM5-X0075Urel330}. %
  All simulations were initialized with the same diameters and at $X=0.075$ and $U_\mathrm{rel}=3.3~\mathrm{m/s}$.} \label{Fig:SingleLiquidCompare}
  \end{figure*}

 The comparison with collisions of the same liquids helps to identify the liquid dominating single process steps of the collision of immiscible liquids.
 Figure~\ref{Fig:OsciallationCompare} exemplarily compares the energy components throughout a G50-SOM5 collision to the energy components of a binary collision of the same liquids, G50-G50 and SOM5-SOM5.
As mentioned before, the local surface energy has a maximum before the kinetic energy reaches its minimum for the G50-SOM5 collision. For the single-liquid collisions of G50-G50 and SOM5-SOM5, the local kinetic energy minimum and surface energy maximum coincide. In the single-liquid case, the local minimum is close to zero, while during the interaction of immiscible droplets almost 10\% of the kinetic energy is remaining.
 The spreading of the phases illustrated in Fig.~\ref{Fig:SingleLiquidCompare} explains this: Two equal droplets spread symmetrically on the collision plane up to a maximum diameter, while already during the spreading phase of the immiscible liquid collision, the outer liquid flows around the inner droplet like around an obstacle. The contraction of the collision complex after its maximum lateral extension is obviously dominated by the encapsulated liquid, as this motion is comparable to the single-liquid collision of two G50 droplets, while the retraction in the pure SOM5 collision case is much slower, cf. Fig.~\ref{Fig:SingleLiquidCompare}. Also the time instant of the kinetic energy minima and maxima of the immiscible droplets and the G50-G50 collision case coincide almost, cf. Fig.~\ref{Fig:OsciallationCompare}. Later, during the elongation of the collision complex along the impact axis, it seems to be the encapsulating liquid that governs the process and prevents another contraction. Instead, the collision of immiscible liquids results in separation, which is a remarkable difference to both single-liquid collisions that still result in coalescence at this velocity and otherwise identical collision setup. In accordance to the different collision outcome, the hotspots of dissipation reveal a different pattern, see Fig.~\ref{Fig:SingleLiquidCompare}. In the collision of same liquids, they are localized symmetrically in the collision plane. The immiscible liquid interaction has those hotspots close to phase boundaries, especially where thin films of liquid or gas, liquid ligaments or triple lines of all three phases are present. \\
 Such analysis on the dominant phases can form the basis for future modeling of the interaction of immiscible liquids in air.
 

%

%% file: 04_Conclusion.tex
\section{Conclusion and Outlook} \label{Sec:Conclusion and Outlook}%

A numerical framework is presented that enables predictive three dimensional Direct Numerical Simulations of the interaction of immiscible fluids, like binary drop collisions. %
The challenges in describing those interactions lie in particular in their multi-scale nature, where local forces at tripe lines and thin films of the two interacting liquids embedded in a surrounding gas phase determine the outcome of the entire process. %
The Volume of Fluid method was extended to include approaches for efficient positioning of the interfaces in three-phase configurations and a robust, energy-conserving calculation of the acceleration due to the interfacial forces, which is based on the Continous Surface Stress model and applicable to thin films, triple lines and large topology changes. %
The presented methods were parallelized with a domain decomposition by means of Message Passing Interface (MPI) and thus enable an efficient computation of arbitrary three dimensional arrangements without the need for local grid refinement. %
With the implementation in the multi-phase solver Free Surface 3D (FS3D), the methods were successfully validated using analytical test cases and comparisons with experimental literature data. \\
The evaluation of the experimentally inaccessible data like the contributions of the kinetic, surface and dissipative energy of both immiscible liquids during the collision process form the basis for its future modeling. %
A comparison of the energy budgets for head-on and off-center coalescence as well as stretching and crossing separation reveals no major differences. %
In contrast to binary collisions of the same liquid, the surface energy maximum occurs slightly earlier than the kinetic energy minimum for all collisions investigated. %
The comparison of the energy budget and also the morphology of the collision of immiscible droplets with the binary collisions of the same liquids, however, is much more revealing. %
While the immiscible collision is initially governed by the motion and time scale of the encapsulated liquid, it is the encapsulating liquid which determines the process at later times. %
This can change the collision outcome. %
For example, crossing separation was observed for an investigated collision of immiscible droplets, while the collisions of droplets of either of the two liquids with an otherwise identical setup resulted both in coalescence. %
Based on these results, future work can identify opportunities to extend existing models on single liquid collisions with respect to their application to immiscible liquids. %

%% file: AppendixLiquidLens.tex
\section{Analytical Solution for the Liquid Lens in a Three Dimensional Setup}\label{App:LiquidLens}
The lens shape is determined by the given surface and interfacial tensions, geometrical considerations and the initialized volume.
If partially wetting liquid combinations are present, the resulting closed Neumann's triangle yields the relations, cf. Fig.~\ref{Fig:LiquidLensSketch} (right),
\begin{equation}
 \begin{aligned}
\cos \theta_{13}= &\frac{\sigma _{23}^2+\sigma _{13}^2-\sigma _{12}^2}{2\sigma _{23}\sigma _{13}} \\
 \cos \theta_{12}=  &\frac{\sigma _{23}^2+\sigma _{12}^2-\sigma _{13}^2}{2\sigma _{12}\sigma _{23}} \text{.} \label{Eq:NeumannTriangle}
 \end{aligned} 
\end{equation}
The base circle radius of the liquid lens, $a$, results from the radius of the spherical caps, $R_{mn}$, and the corresponding contact angles, $\theta_{mn}$, according to
\begin{equation}
a = R_{mn}\sin(\theta_{mn})\text{.}
\end{equation}
The radius $R_{mn}$ results from the volume balance 
\begin{equation}
V_{0,1} =  V_{SC,12}  +   V_{SC,13} 
\end{equation}
with the initial volume of liquid~$1$
\begin{equation}
V_{0,1} = \frac{4\pi}{3} R_{0,1}^3
\end{equation}
and the volume of each spherical cap,
\begin{equation}
V_{\mathrm{SC},mn} =  \frac{\pi}{3}  R_{mn}^3 \left(2+\cos\left(\theta_{mn}\right)\right)\left(1-\cos\left(\theta_{mn}\right)\right)^2 \text{.}
\end{equation}
This leads to
\begin{equation}
a =  R_{0,1} \left( \frac{4}{\phi_{12} + \phi_{13}} \right)^\frac{1}{3} 
\end{equation}
with
\begin{equation}
\phi_{mn} =  \frac{1}{\left(\sin\left(\theta_{mn}\right)\right)^3} \left(2+\cos\left(\theta_{mn}\right)\right)\left(1-\cos\left(\theta_{mn}\right)\right)^2 \text{.}
\end{equation}
The height of each spherical cap is given by
\begin{equation}
h_{mn}   = a \tan{\frac{\theta_{mn}}{2}}
\end{equation}
by geometric relations.

%% file: AppendixEnergy.tex
\section{Energy Evaluation}\label{App:Energy}
The details of the evaluation of the energy contributions and the viscous dissipation rates are presented in the following. %
\subsection{Kinetic Energy} \label{subsec:KineticEnergy}
The kinetic energy %
\begin{equation}
E_\mathrm{kin} =  \frac{1}{2} (\Delta x)^3 \rho_m \sum_{m=1}^3 \sum_{\mathrm{cell}=1}^{N_\mathrm{cells}} f_{m,\mathrm{cell}} (|\mathbf{u}|^2)_\mathrm{cell}
\label{Eq:Ekin}
\end{equation}
was evaluated with 
\begin{equation}
\begin{aligned}
  (|\mathbf{u}|^2)_\mathrm{cell} =  \frac{1}{3}& \left( u_{i-}^2 + u_{i-} u_{i+} + u_{i+}^2 \right.  \\
        &\left.+  v_{j-}^2 + v_{j-} v_{j+} + v_{j+}^2\right.  \\
        &\left.+  w_{k-}^2 + w_{k-} w_{k+} + w_{k+}^2 \right) \\
      \label{Eq:usqared_staggered}
  \end{aligned}
\end{equation}
to obtain values in the center of the cell $(i,j,k)$ in the staggered grid arrangement. %
The indices $+$ and $-$ indicate the left and right faces of the cell, where the velocity components are stored in the staggered grid. %
\subsection{Surface Energy}
The surface energy %
\begin{equation}
 E_\sigma = (\Delta x)^3 \sum_{m=1}^3 \gamma_m  \sum_{\mathrm{cell}=1}^{N_\mathrm{cells}} |\nabla f_{m,\mathrm{cell}}|
 \label{Eq:Esigma}
\end{equation}
was obtained from superposition of the contributions of each phase with the aid of the partial surface tensions $\gamma_m$ from Eq.~\ref{Eq:SigmaDecompose}. %
\subsection{Dissipated Energy and Local Viscous Dissipation Rate}\label{subsec:DissipatedEnergy} %
Evaluating the dissipation function, 
\begin{equation}
\begin{aligned}
\Phi(x,y,z,t) = \mu \Bigg[&   2 \left(\frac{\partial u(x,y,z,t)}{\partial x}\right)^2 \\
+ &   2 \left(\frac{\partial v(x,y,z,t)}{\partial y}\right)^2 \\
+&  2 \left(\frac{\partial w(x,y,z,t)}{\partial z}\right)^2 \\
+ &   \left(\frac{\partial v(x,y,z,t)}{\partial x} + \frac{\partial u(x,y,z,t)}{\partial y}\right)^2 \\
+ &   \left(\frac{\partial w(x,y,z,t)}{\partial y} + \frac{\partial v(x,y,z,t)}{\partial z}\right)^2  \\
+&  \left(\frac{\partial u(x,y,z,t)}{\partial z} + \frac{\partial w(x,y,z,t)}{\partial x}\right)^2  \Bigg] \text{,}\\
\end{aligned} \label{Eq:VDRl}
\end{equation}
in a discrete cell yields the Viscous Dissipation Rate, VDR, used throughout this paper. A Newtonian fluid is assumed, and the volume viscosity is neglected \citep{White1991}.
The volume and time integral of the local dissipation function yields the dissipated energy until time $t_n$  in the semi-discrete form %
\begin{equation}
E_\mathrm{diss}(t_n) =  \int_0^{t_n} \left( (\Delta x)^3 \sum_{\mathrm{cell}=1}^{N_\mathrm{cells}} \Phi_\mathrm{cell}(t)\right)~dt \text{.}
\end{equation}
The discrete form of the dissipated energy reads
\begin{equation}
 E_\mathrm{diss}(t_n) =  \sum_{n_t=0}^{N_t} \left((\Delta x)^3 \sum_{\mathrm{cell}=1}^{N_\mathrm{cells}} \Phi_\mathrm{cell,n_t}\right)\Delta t_{n_t} 
\label{Eq:Ediss}
\end{equation}
with the cell-mean local viscous dissipation rate $\Phi_\mathrm{cell,n_t}$. It represents the evaluation of $\Phi(x,y,z,t)$ at the $n_t$-th time step in a discrete cell $i,j,k$ of dimension $(\Delta x )^3$ in the Cartesian mesh with a total number of $N_\mathrm{cells}$. The time-step varies according to the Courant-Friedrichs-Lewy (CFL) condition.  %
The spatial derivatives required for the computation of $\Phi_\mathrm{cell}$ are approximated with first order central differences. %
It should be noted, that the central differences utilized for the approximation of the velocity gradients for the "off-diagonal" derivatives have to be evaluated over the distance of two cells due to the staggered grid arrangement. %
Thus the evaluation of the viscous shear losses is always performed with a lower resolution than the "on-diagonal" derivatives. %

%% file: AppendixConvergence.tex
\section{Details on the Simulations and Grid Convergence}\label{App:Convergence}
\input{02_01_02_Advection}
\subsection{Energy Conservation and Grid Resolution}\label{App:ConvergenceEnergy}
\begin{figure}
 \centering
\includegraphics[scale=0.7]{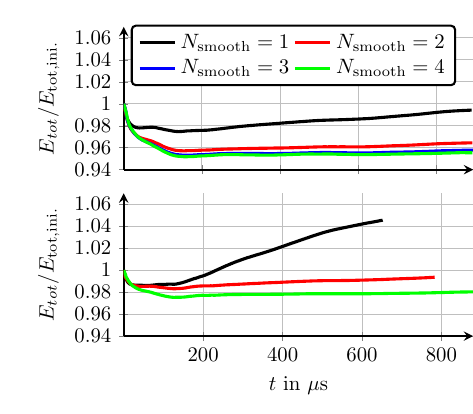}
 \caption{Influence of the employed number of smoothing steps on the energy conservation for an exemplary G50-SOM5 collision with a resolution of $512~{\mathrm{cells}}/{0.8~\mathrm{mm}}$ (top) and $1024~{\mathrm{cells}}/{0.8~\mathrm{mm}}$ (bottom), $X=0.023$, $U_\mathrm{rel} = 3.17~\mathrm{{m}/{s}}$. %
The energy gain with only one smoothing step is significant for both resolutions, with two smoothing steps still noticeable and with four smoothing steps good energy conservation is obtained. %
At the higher resolution the energy gain gets more pronounced. %
The simulations where stopped as soon as the energy gain was clearly visible in order to save computational resources. %
}%
\label{Fig:SmoothingStudy}
 \end{figure}

A potential energy gain caused by the smoothing of the volume fraction field required for the computation of the surfaces forces was discussed in Sec.~\ref{Subsec:SurfaceForces}. Studying the influence of the number of smoothing steps performed with the presented linear smoothing operator on energy conservation, we found that four smoothing steps are leading to energy conserving solutions, while one smoothing step results in significant energy gain. Two and three smoothing steps also yield some energy gain throughout the simulated collisions. The results of this study are shown in Fig.~\ref{Fig:SmoothingStudy}. %
It also is clearly visible, that the energy gain gets more pronounced at higher resolutions. %
The unphysical energy gain leads to separation of the collision complex at lower relative velocities of the droplets and consequently to a shift of the regime boundary. %
Thus, the numerical parameter $N_\mathrm{smooth}$ was set to 4 for the simulations presented in this paper. Only, Perfluo-G50 collisions were performed with $N_\mathrm{smooth}=3$, which yielded better results due to the limited resolution of the ligament. \\

\begin{figure*}
 \centering
\includegraphics[scale=0.7]{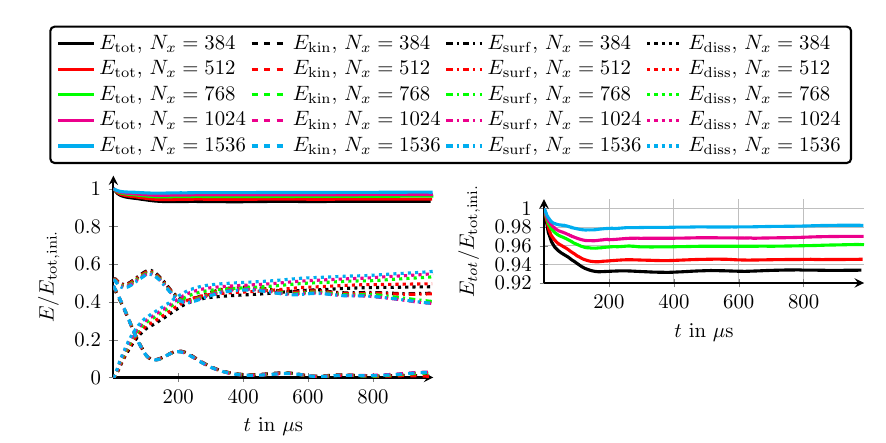}
 \caption{Convergence study of the energy contributions  of G50-SOM5 droplet collisions just below the head-on separation threshold with $U_\mathrm{rel} = 3.2~\mathrm{\frac{m}{s}}$, $X=0$, $N_\mathrm{smooth}=4$. %
A zoom on total energy budget throughout the collision is shown on the right. A resolution of at least $768~ \mathrm{ cells/mm}$ is required for this case  to yield an acceptable total energy loss of below $4\%$. The resolution also influences the collision outcome. For higher resolutions, coalescence is achieved, while the simulations at lower resolutions predict crossing separation. This is indicated by the surface energy, which finally decreases for higher resolutions. 
Thus, a resolution of  $768~ \mathrm{ cells/mm}$ was chosen for the computation of the regime map in Fig.~\ref{Fig:RegimeMapSOM5G50} (or, if the size of the collision complex allowed shrinking the domain size, the slightly higher resolution of $768~\mathrm{cells/0.8~mm} = 960~\mathrm{cells/mm}$).}
 \label{Fig:ConvergenceEnergy}
 \end{figure*}

Figure~\ref{Fig:SmoothingStudy} also shows that, depending on the resolution, the total energy decreases slightly at the beginning of the collision. This is further illustrated in Fig.~\ref{Fig:ConvergenceEnergy} with the representation of the normalized total energy for different resolutions for the exemplary case of a G50-SOM5 droplet collision just below the head-on regime boundary. At the beginning, a thin gas film between the approaching droplets is present, which cannot be resolved.
However, the impact of this under-resolved gas film on the regime boundaries and morphology for all investigated collisions is negligible as the excellent comparison to experiments reveals. \\
Energy conservation is guaranteed for $t \gtrsim 200~\mathrm{\mu s}$ and resolutions above $512~\mathrm{\frac{cells}{mm}}$. For these resolutions, also the critical velocity at the head-on regime boundary converges well. \\
While the kinetic and surface energy are converged well already for the lower resolutions, cf. Fig.~\ref{Fig:ConvergenceEnergy}, the energy of dissipation is the major source of unaccounted energy loss due its discretization, that requires the evaluation of second derivatives in space, cf. \ref{App:Energy}. \\
The resolutions used for the simulations of different liquid combinations in this study were chosen such that the losses attributed to the initial phase make up less than $4\%$ of the whole energy budget and no significant change in the critical velocity at the head-on regime boundary is observed. \\

\subsection{Computational Infrastructure}\label{App:ComputationalInfrastructure}
All simulations ran on the supercomputer HPE Apollo (Hawk) of the High Performance Computing Center of the University of Stuttgart (HLRS). %
The cases from the validation in Subsec.~\ref{Subsec:ExperimentCompare} with $768\times192$ or $384\times192$ cells in a $0.8\times0.2$, $0.4\times0.2~\mathrm{mm}$ or $1\times0.25\times0.25 \mathrm{mm}$ domain, or $1024\times512\times256$ in a $1.2\times0.6\times0.3~\mathrm{mm}$ domain ran for around 70 to 90~h wall-time on the supercomputer. The three dimensional DNS ran on 512 (exactly head-on cases, two symmetry planes) or 1024 CPUs (off-centre cases, only one symmetry plane). %
Performance tests show, that FS3D scales well up to 1024 cores \citep{HLRSBericht2022Preprint}, which is valid also for the simulation of immiscible liquids' interaction. %
The $1536\times384\times384$-cells cases for the convergence study shown in \ref{App:ConvergenceEnergy} ran for 11 to 12 days on 1024 CPUs, an infeasible runtime for a larger parameter study. %
Increasing both the single CPU's performance and the parallel scaling in order to increase the feasible resolutions and domain sizes within a reasonable compute time is currently work in progress.%


%% file: 02_01_02_Advection.tex
\subsection{Advection Test} \label{App:AdvectTest} %
\begin{figure}
\centering
\includegraphics[width=2cm, trim = 30cm 4cm 30cm 4cm, clip]{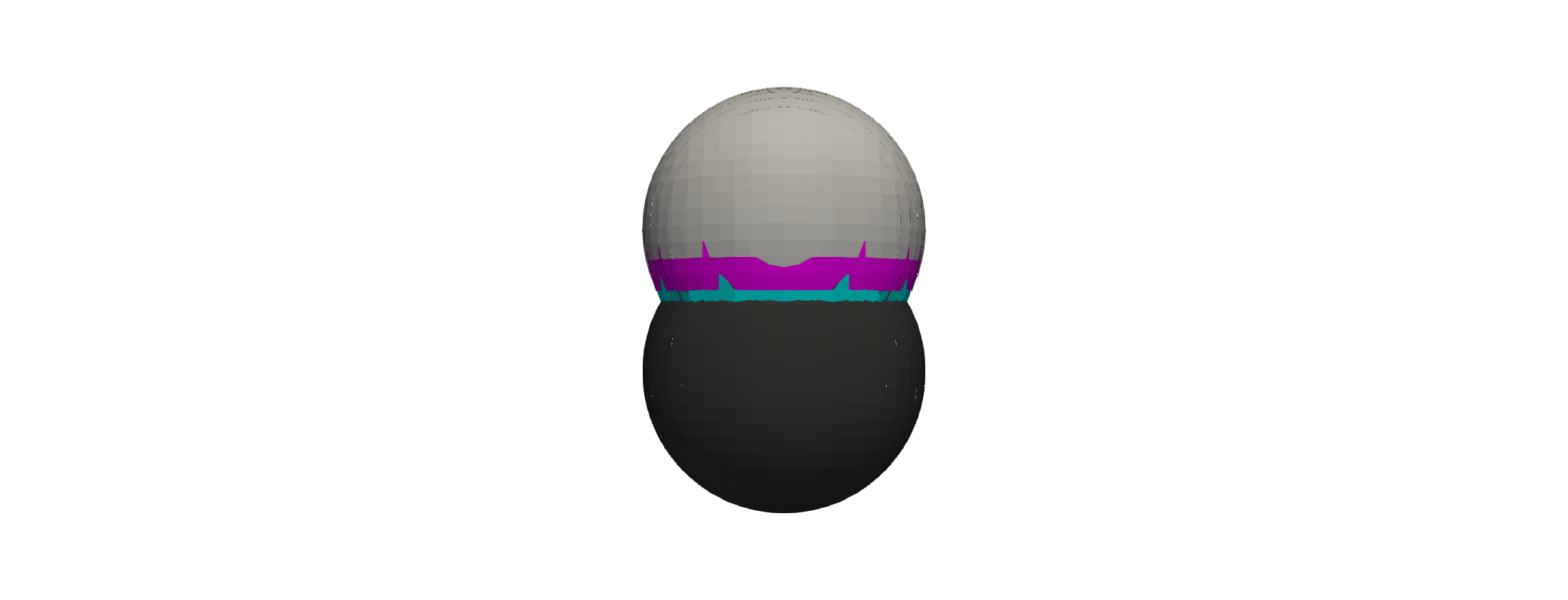}
\caption{Prototypical collision complex of a sphere of liquid~1 (dark grey) and a spherical cap of liquid~2 (light grey) as a representative advection test case, $\theta_\text{ini}=60^\circ$, $\mathrm{sphere}=3~\mathrm{mm}$, exemplary resolved with $64~\mathrm{cells}/10~\mathrm{mm} = 19.2~\mathrm{cells}/D_\mathrm{sphere}$.  %
The liquid~$2$'s interfaces reconstructed with the three-phase PLIC algorithm are marked according to Figs.~\ref{Fig:3phTopology} and \ref{Fig:ImmiscibleExample}.
}\label{Fig:Case3Theta60deg}
\end{figure}

\begin{figure}
\centering
\includegraphics[scale=1]{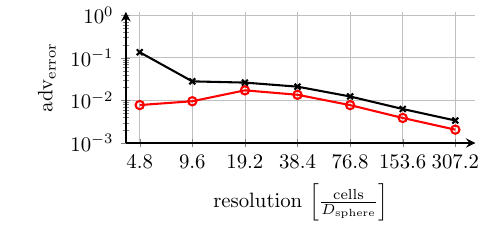}
\caption{Errors $\mathrm{adv}_{\mathrm{error}}$ of the volume fractions $f_{1}$ (red) and $f_{2}$ (black) after advection for exemplary test case (cf. Fig.~\ref{Fig:Case3Theta60deg}) are shown for different grid resolutions.}
\label{Fig:ConvergenceAdvection}
\end{figure}

The advection and reconstruction of the volume fractions within three-phase cells was validated by translating a sphere of liquid~1 with a spherical cap of liquid~2 residing on top, cf. Fig.~\ref{Fig:Case3Theta60deg}, diagonally through a cuboidal domain with periodic boundaries and evaluating the deviation of the shape in terms of volume fractions after passing the original position again. A contact angle $\theta_\text{ini}$ is prescribed by the spheres' intersection resulting from the given distance of the center points. It is set to $\theta_\mathrm{ini} = 60^\circ$ in this representative test case. %
The diameters of both spheres are equal, $D_\mathrm{sphere}=D_1=D_2=3~\mathrm{mm}$, initialized within a $(10~\mathrm{mm})^3$ domain. %
The interfaces of liquid~2 (light grey) which are in the same cell like liquid~1 (dark grey) are marked with cyan, those that are in cells lying inside the reconstruction stencil of liquid~1 are marked in magenta. The difference in the average volume fraction after advection
\begin{equation}
\mathrm{adv}_{\mathrm{error}} = \frac{\sum\limits_{\mathrm{cell}=1}^{N_\mathrm{cells}}| f_{m,\mathrm{ini},\mathrm{cell}}-f_{m,\mathrm{cell}}|}{\sum\limits_{\mathrm{cell}=1}^{N_\mathrm{cells}} f_{m,\mathrm{ini},\mathrm{cell}}} \text{,}
\end{equation} 
with $N_\mathrm{cells}$ the total number of cells in the computational domain, $f_{m,\text{ini},\text{cell}}$ the initialized volume fraction of the respective liquid and $f_{m,\text{cell}}$, the volume fraction after the advection, is shown in Fig.~\ref{Fig:ConvergenceAdvection} for different grid resolutions,  %
It should be noted, that the error for the reconstruction of the interface of liquid~2 is always higher than the one of liquid~1 due to the sequential PLIC reconstruction of liquid~2 interfaces upon the reconstructed liquid~1 interfaces. %
An experimental order of convergence around 1 is reached, which is of the same order as the original two-phase PLIC algorithm. %

%% file: PotykaSchulte2023_arXivV2_ImmiscibleInteractionVOF.bbl
\begin{thebibliography}{74}
\expandafter\ifx\csname natexlab\endcsname\relax\def\natexlab#1{#1}\fi
\providecommand{\url}[1]{\texttt{#1}}
\providecommand{\href}[2]{#2}
\providecommand{\path}[1]{#1}
\providecommand{\DOIprefix}{doi:}
\providecommand{\ArXivprefix}{arXiv:}
\providecommand{\URLprefix}{URL: }
\providecommand{\Pubmedprefix}{pmid:}
\providecommand{\doi}[1]{\href{http://dx.doi.org/#1}{\path{#1}}}
\providecommand{\Pubmed}[1]{\href{pmid:#1}{\path{#1}}}
\providecommand{\bibinfo}[2]{#2}
\ifx\xfnm\relax \def\xfnm[#1]{\unskip,\space#1}\fi
\bibitem[{Ashgriz and Poo(1990)}]{Ashgriz1990}
\bibinfo{author}{Ashgriz, N.}, \bibinfo{author}{Poo, J.Y.},
  \bibinfo{year}{1990}.
\newblock \bibinfo{title}{Coalescence and separation in binary collisions of
  liquid drops}.
\newblock \bibinfo{journal}{Journal of Fluid Mechanics} \bibinfo{volume}{221},
  \bibinfo{pages}{183 -- 204}.
\newblock \DOIprefix\doi{10.1017/S0022112090003536}.
\bibitem[{Baumgartner et~al.(2022)Baumgartner, Brenn and
  Planchette}]{Baumgartner2022}
\bibinfo{author}{Baumgartner, D.}, \bibinfo{author}{Brenn, G.},
  \bibinfo{author}{Planchette, C.}, \bibinfo{year}{2022}.
\newblock \bibinfo{title}{Universality of stretching separation}.
\newblock \bibinfo{journal}{Journal of Fluid Mechanics} \bibinfo{volume}{937},
  \bibinfo{pages}{R1}.
\newblock \DOIprefix\doi{10.1017/jfm.2022.107}.
\bibitem[{Benson(2002)}]{Benson2002}
\bibinfo{author}{Benson, D.J.}, \bibinfo{year}{2002}.
\newblock \bibinfo{title}{{Volume of fluid interface reconstruction methods for
  multi-material problems}}.
\newblock \bibinfo{journal}{Applied Mechanics Reviews} \bibinfo{volume}{55},
  \bibinfo{pages}{151--165}.
\newblock \DOIprefix\doi{10.1115/1.1448524}.
\bibitem[{Brackbill et~al.(1992)Brackbill, Kothe and Zemach}]{Brackbill1992}
\bibinfo{author}{Brackbill, J.U.}, \bibinfo{author}{Kothe, D.B.},
  \bibinfo{author}{Zemach, C.}, \bibinfo{year}{1992}.
\newblock \bibinfo{title}{{A Continuum Method for Modeling Surface Tension}}.
\newblock \bibinfo{journal}{Journal of Computational Physics}
  \bibinfo{volume}{100}, \bibinfo{pages}{335--354}.
\newblock \DOIprefix\doi{10.1016/0021-9991(92)90240-Y}.
\bibitem[{Chen and Chen(2006)}]{Chen2006a}
\bibinfo{author}{Chen, R.H.}, \bibinfo{author}{Chen, C.T.},
  \bibinfo{year}{2006}.
\newblock \bibinfo{title}{Collision between immiscible drops with large surface
  tension difference: Diesel oil and water}.
\newblock \bibinfo{journal}{Experiments in Fluids} \bibinfo{volume}{41},
  \bibinfo{pages}{453--461}.
\newblock \DOIprefix\doi{10.1007/s00348-006-0173-2}.
\bibitem[{Chen and Yang(2020)}]{Chen2020}
\bibinfo{author}{Chen, X.}, \bibinfo{author}{Yang, V.}, \bibinfo{year}{2020}.
\newblock \bibinfo{title}{Direct numerical simulation of multiscale flow
  physics of binary droplet collision}.
\newblock \bibinfo{journal}{Physics of Fluids} \bibinfo{volume}{32},
  \bibinfo{pages}{062103}.
\newblock \DOIprefix\doi{10.1063/5.0006695}.
\bibitem[{Ebadi and Hosseinalipour(2022)}]{Ebadi2022}
\bibinfo{author}{Ebadi, A.}, \bibinfo{author}{Hosseinalipour, S.M.},
  \bibinfo{year}{2022}.
\newblock \bibinfo{title}{The collision of immiscible droplets in three-phase
  liquid systems: A numerical study using phase-field lattice boltzmann
  method}.
\newblock \bibinfo{journal}{Chemical Engineering Research and Design}
  \bibinfo{volume}{178}, \bibinfo{pages}{289--314}.
\newblock \DOIprefix\doi{10.1016/j.cherd.2021.12.019}.
\bibitem[{Eisenschmidt et~al.(2016)Eisenschmidt, Ertl, Gomaa, Kieffer-Roth,
  Meister, Rauschenberger, Reitzle, Schlottke and Weigand}]{Eisenschmidt2016}
\bibinfo{author}{Eisenschmidt, K.}, \bibinfo{author}{Ertl, M.},
  \bibinfo{author}{Gomaa, H.}, \bibinfo{author}{Kieffer-Roth, C.},
  \bibinfo{author}{Meister, C.}, \bibinfo{author}{Rauschenberger, P.},
  \bibinfo{author}{Reitzle, M.}, \bibinfo{author}{Schlottke, K.},
  \bibinfo{author}{Weigand, B.}, \bibinfo{year}{2016}.
\newblock \bibinfo{title}{Direct numerical simulations for multiphase flows: An
  overview of the multiphase code {FS3D}}.
\newblock \bibinfo{journal}{Applied Mathematics and Computation}
  \bibinfo{volume}{272}, \bibinfo{pages}{508--517}.
\newblock \DOIprefix\doi{10.1016/j.amc.2015.05.095}.
\bibitem[{Finotello et~al.(2017)Finotello, Padding, Deen, Jongsma, Innings and
  Kuipers}]{Finotello2017}
\bibinfo{author}{Finotello, G.}, \bibinfo{author}{Padding, J.T.},
  \bibinfo{author}{Deen, N.G.}, \bibinfo{author}{Jongsma, A.},
  \bibinfo{author}{Innings, F.}, \bibinfo{author}{Kuipers, J.A.M.},
  \bibinfo{year}{2017}.
\newblock \bibinfo{title}{Effect of viscosity on droplet-droplet collisional
  interaction}.
\newblock \bibinfo{journal}{Physics of Fluids} \bibinfo{volume}{29},
  \bibinfo{pages}{067102}.
\newblock \DOIprefix\doi{10.1063/1.4984081}.
\bibitem[{Gotaas et~al.(2007)Gotaas, Havelka, Jakobsen, Svendsen, Hase, Roth
  and Weigand}]{Gotaas2007}
\bibinfo{author}{Gotaas, C.}, \bibinfo{author}{Havelka, P.},
  \bibinfo{author}{Jakobsen, H.A.}, \bibinfo{author}{Svendsen, H.F.},
  \bibinfo{author}{Hase, M.}, \bibinfo{author}{Roth, N.},
  \bibinfo{author}{Weigand, B.}, \bibinfo{year}{2007}.
\newblock \bibinfo{title}{{Effect of viscosity on droplet-droplet collision
  outcome: Experimental study and numerical comparison}}.
\newblock \bibinfo{journal}{Physics of Fluids} \bibinfo{volume}{19},
  \bibinfo{pages}{102106}.
\newblock \DOIprefix\doi{10.1063/1.2781603}.
\bibitem[{Haghani Hassan~Abadi et~al.(2018)Haghani Hassan~Abadi, Fakhari and
  Rahimian}]{HaghaniHassanAbadi2018}
\bibinfo{author}{Haghani Hassan~Abadi, R.}, \bibinfo{author}{Fakhari, A.},
  \bibinfo{author}{Rahimian, M.H.}, \bibinfo{year}{2018}.
\newblock \bibinfo{title}{Numerical simulation of three-component multiphase
  flows at high density and viscosity ratios using lattice boltzmann methods}.
\newblock \bibinfo{journal}{Phys. Rev. E} \bibinfo{volume}{97},
  \bibinfo{pages}{033312}.
\newblock \DOIprefix\doi{10.1103/PhysRevE.97.033312}.
\bibitem[{Harlow and Welch(1965)}]{Harlow1965}
\bibinfo{author}{Harlow, F.}, \bibinfo{author}{Welch, J.},
  \bibinfo{year}{1965}.
\newblock \bibinfo{title}{{Numerical calculation of time-dependent viscous
  incompressible flow of fluid with free surface}}.
\newblock \bibinfo{journal}{Physics of Fluids} \bibinfo{volume}{8},
  \bibinfo{pages}{2182--2189}.
\newblock \DOIprefix\doi{10.1063/1.1761178}.
\bibitem[{He et~al.(2019)He, Xia and Zhang}]{He2019}
\bibinfo{author}{He, C.}, \bibinfo{author}{Xia, X.}, \bibinfo{author}{Zhang,
  P.}, \bibinfo{year}{2019}.
\newblock \bibinfo{title}{Non-monotonic viscous dissipation of bouncing
  droplets undergoing off-center collision}.
\newblock \bibinfo{journal}{Physics of Fluids} \bibinfo{volume}{31},
  \bibinfo{pages}{052004}.
\newblock \DOIprefix\doi{10.1063/1.5088544}.
\bibitem[{He et~al.(2020)He, Xia and Zhang}]{He2020}
\bibinfo{author}{He, C.}, \bibinfo{author}{Xia, X.}, \bibinfo{author}{Zhang,
  P.}, \bibinfo{year}{2020}.
\newblock \bibinfo{title}{Vortex-dynamical implications of nonmonotonic viscous
  dissipation of off-center droplet bouncing}.
\newblock \bibinfo{journal}{Physics of Fluids} \bibinfo{volume}{32},
  \bibinfo{pages}{032004}.
\newblock \DOIprefix\doi{10.1063/5.0003057}.
\bibitem[{Hirt and Nichols(1981)}]{Hirt1981}
\bibinfo{author}{Hirt, C.}, \bibinfo{author}{Nichols, B.},
  \bibinfo{year}{1981}.
\newblock \bibinfo{title}{{Volume of fluid {(VOF)} method for the dynamics of
  free boundaries}}.
\newblock \bibinfo{journal}{Journal of Computational Physics}
  \bibinfo{volume}{39 (1)}, \bibinfo{pages}{201--225}.
\newblock \DOIprefix\doi{10.1016/0021-9991(81)90145-5}.
\bibitem[{Ho and Vu(2023)}]{Ho2023}
\bibinfo{author}{Ho, N.X.}, \bibinfo{author}{Vu, T.V.}, \bibinfo{year}{2023}.
\newblock \bibinfo{title}{{Numerical study of head-on collision of two
  equal-sized compound droplets}}.
\newblock \bibinfo{journal}{Physics of Fluids} \bibinfo{volume}{35},
  \bibinfo{pages}{063320}.
\newblock \DOIprefix\doi{10.1063/5.0153227}.
\bibitem[{Inamuro et~al.(2004)Inamuro, Tajima and Ogino}]{Inamuro2004}
\bibinfo{author}{Inamuro, T.}, \bibinfo{author}{Tajima, S.},
  \bibinfo{author}{Ogino, F.}, \bibinfo{year}{2004}.
\newblock \bibinfo{title}{{Lattice Boltzmann simulation of droplet collision
  dynamics}}.
\newblock \bibinfo{journal}{International Journal of Heat Mass Transfer}
  \bibinfo{volume}{47}, \bibinfo{pages}{4649--4657}.
\newblock \DOIprefix\doi{10.1016/j.ijheatmasstransfer.2003.08.030}.
\bibitem[{Jiang et~al.(1992)Jiang, Umemura and Law}]{Jiang1992}
\bibinfo{author}{Jiang, Y.J.}, \bibinfo{author}{Umemura, A.},
  \bibinfo{author}{Law, C.K.}, \bibinfo{year}{1992}.
\newblock \bibinfo{title}{An experimental investigation on the collision
  behaviour of hydrocarbon droplets}.
\newblock \bibinfo{journal}{Journal of Fluid Mechanics} \bibinfo{volume}{234},
  \bibinfo{pages}{171–190}.
\newblock \DOIprefix\doi{10.1017/S0022112092000740}.
\bibitem[{Joubert et~al.(2020)Joubert, Gardin, Zaleski and
  Popinet}]{Joubert2020}
\bibinfo{author}{Joubert, N.}, \bibinfo{author}{Gardin, P.},
  \bibinfo{author}{Zaleski, S.}, \bibinfo{author}{Popinet, S.},
  \bibinfo{year}{2020}.
\newblock \bibinfo{title}{Modelling of mass transfer in a steelmaking ladle},
  in: \bibinfo{booktitle}{14th International Conference on CFD in Oil \& Gas,
  Metallurgical and Process Industries}.
\bibitem[{Kamperman et~al.(2018)Kamperman, Trikalitis, Karperien, Visser and
  Leijten}]{Kamperman2018}
\bibinfo{author}{Kamperman, T.}, \bibinfo{author}{Trikalitis, V.D.},
  \bibinfo{author}{Karperien, M.}, \bibinfo{author}{Visser, C.W.},
  \bibinfo{author}{Leijten, J.}, \bibinfo{year}{2018}.
\newblock \bibinfo{title}{Ultrahigh-throughput production of monodisperse and
  multifunctional janus microparticles using in-air microfluidics}.
\newblock \bibinfo{journal}{ACS Applied Materials \& Interfaces}
  \bibinfo{volume}{10}, \bibinfo{pages}{23433--23438}.
\newblock \DOIprefix\doi{10.1021/acsami.8b05227}.
\bibitem[{Kromer et~al.(2023)Kromer, Potyka, Schulte and
  Bothe}]{Kromer2021ArXiv}
\bibinfo{author}{Kromer, J.}, \bibinfo{author}{Potyka, J.},
  \bibinfo{author}{Schulte, K.}, \bibinfo{author}{Bothe, D.},
  \bibinfo{year}{2023}.
\newblock \bibinfo{title}{{Efficient sequential PLIC interface positioning for
  enhanced performance of the three-phase VoF Method (accepted with minor
  revisons at Computers and Fluids, arXiv preprint)}}.
\newblock \DOIprefix\doi{10.48550/ARXIV.2105.08972}.
\bibitem[{Lafaurie et~al.(1994)Lafaurie, Nardone, Scardovelli, Zaleski and
  Zanetti}]{Lafaurie1994}
\bibinfo{author}{Lafaurie, B.}, \bibinfo{author}{Nardone, C.},
  \bibinfo{author}{Scardovelli, R.}, \bibinfo{author}{Zaleski, S.},
  \bibinfo{author}{Zanetti, G.}, \bibinfo{year}{1994}.
\newblock \bibinfo{title}{{Modelling Merging and Fragmentation in Multiphase
  Flows with {SURFER}}}.
\newblock \bibinfo{journal}{Journal of Computational Physics}
  \bibinfo{volume}{113}, \bibinfo{pages}{134--147}.
\newblock \DOIprefix\doi{10.1006/jcph.1994.1123}.
\bibitem[{Li et~al.(2015)Li, Guo, Jemison, Sussman, Helms and Arienti}]{Li2015}
\bibinfo{author}{Li, G.and~Lian, Y.}, \bibinfo{author}{Guo, Y.},
  \bibinfo{author}{Jemison, M.}, \bibinfo{author}{Sussman, M.},
  \bibinfo{author}{Helms, T.}, \bibinfo{author}{Arienti, M.},
  \bibinfo{year}{2015}.
\newblock \bibinfo{title}{Incompressible multiphase flow and encapsulation
  simulations using the moment-of-fluid method}.
\newblock \bibinfo{journal}{International Journal for Numerical Methods in
  Fluids} \bibinfo{volume}{79}, \bibinfo{pages}{456--490}.
\newblock \DOIprefix\doi{10.1002/fld.4062}.
\bibitem[{Liu and Bothe(2016)}]{Liu2016}
\bibinfo{author}{Liu, M.}, \bibinfo{author}{Bothe, D.}, \bibinfo{year}{2016}.
\newblock \bibinfo{title}{Numerical study of head-on droplet collisions at high
  weber numbers}.
\newblock \bibinfo{journal}{Journal of Fluid Mechanics} \bibinfo{volume}{789},
  \bibinfo{pages}{785--805}.
\newblock \DOIprefix\doi{10.1017/jfm.2015.725}.
\bibitem[{Lycett-Brown et~al.(2014)Lycett-Brown, Luo, Liu and
  Lv}]{LycettBrown2014}
\bibinfo{author}{Lycett-Brown, D.}, \bibinfo{author}{Luo, K.H.},
  \bibinfo{author}{Liu, R.}, \bibinfo{author}{Lv, P.}, \bibinfo{year}{2014}.
\newblock \bibinfo{title}{{Binary droplet collision simulations by a multiphase
  cascaded lattice Boltzmann method}}.
\newblock \bibinfo{journal}{Physics of Fluids} \bibinfo{volume}{26},
  \bibinfo{pages}{023303}.
\newblock \DOIprefix\doi{10.1063/1.4866146}.
\bibitem[{Mazloomi~Moqaddam et~al.(2016)Mazloomi~Moqaddam, Chikatamarla and
  Karlin}]{Mazloomi2016}
\bibinfo{author}{Mazloomi~Moqaddam, A.}, \bibinfo{author}{Chikatamarla, S.S.},
  \bibinfo{author}{Karlin, I.V.}, \bibinfo{year}{2016}.
\newblock \bibinfo{title}{{Simulation of binary droplet collisions with the
  entropic lattice Boltzmann method}}.
\newblock \bibinfo{journal}{Physics of Fluids} \bibinfo{volume}{28},
  \bibinfo{pages}{022106}.
\newblock \DOIprefix\doi{10.1063/1.4942017}.
\bibitem[{Munnannur and Reitz(2007)}]{Munnannur2007}
\bibinfo{author}{Munnannur, A.}, \bibinfo{author}{Reitz, R.D.},
  \bibinfo{year}{2007}.
\newblock \bibinfo{title}{A new predictive model for fragmenting and
  non-fragmenting binary droplet collisions}.
\newblock \bibinfo{journal}{International Journal of Multiphase Flow}
  \bibinfo{volume}{33}, \bibinfo{pages}{873--896}.
\newblock \DOIprefix\doi{10.1016/j.ijmultiphaseflow.2007.03.003}.
\bibitem[{Nikolopoulos et~al.(2009a)Nikolopoulos, Nikas and
  Bergeles}]{Nikolopoulos2009}
\bibinfo{author}{Nikolopoulos, N.}, \bibinfo{author}{Nikas, K.S.},
  \bibinfo{author}{Bergeles, G.}, \bibinfo{year}{2009}a.
\newblock \bibinfo{title}{A numerical investigation of central binary collision
  of droplets}.
\newblock \bibinfo{journal}{Computers \& Fluids} \bibinfo{volume}{38},
  \bibinfo{pages}{1191--1202}.
\newblock \DOIprefix\doi{10.1016/j.compfluid.2008.11.007}.
\bibitem[{Nikolopoulos et~al.(2009b)Nikolopoulos, Theodorakakos and
  Bergeles}]{Nikolopoulos2009a}
\bibinfo{author}{Nikolopoulos, N.}, \bibinfo{author}{Theodorakakos, A.},
  \bibinfo{author}{Bergeles, G.}, \bibinfo{year}{2009}b.
\newblock \bibinfo{title}{Off-centre binary collision of droplets: A numerical
  investigation}.
\newblock \bibinfo{journal}{International Journal of Heat Mass Transfer}
  \bibinfo{volume}{52}, \bibinfo{pages}{4160--4174}.
\newblock \DOIprefix\doi{10.1016/j.ijheatmasstransfer.2009.04.011}.
\bibitem[{Nobari and Tryggvason(1996)}]{Nobari1996a}
\bibinfo{author}{Nobari, M.}, \bibinfo{author}{Tryggvason, G.},
  \bibinfo{year}{1996}.
\newblock \bibinfo{title}{Numerical simulations of three-dimensional drop
  collisions}.
\newblock \bibinfo{journal}{AIAA J.} \bibinfo{volume}{34},
  \bibinfo{pages}{750--755}.
\newblock \DOIprefix\doi{10.2514/3.13136}.
\bibitem[{Nobari et~al.(1996)Nobari, Jan and Tryggvason}]{Nobari1996}
\bibinfo{author}{Nobari, M.R.}, \bibinfo{author}{Jan, Y.J.},
  \bibinfo{author}{Tryggvason, G.}, \bibinfo{year}{1996}.
\newblock \bibinfo{title}{{Head-on collision of drops-A numerical
  investigation}}.
\newblock \bibinfo{journal}{Physics of Fluids} \bibinfo{volume}{8},
  \bibinfo{pages}{29--42}.
\newblock \DOIprefix\doi{10.1063/1.868812}.
\bibitem[{Orme(1997)}]{Orme1997}
\bibinfo{author}{Orme, M.}, \bibinfo{year}{1997}.
\newblock \bibinfo{title}{Experiments on droplet collisions, bounce,
  coalescence and disruption}.
\newblock \bibinfo{journal}{Progress in Energy and Combustion Science}
  \bibinfo{volume}{23}, \bibinfo{pages}{65--79}.
\newblock \DOIprefix\doi{10.1016/S0360-1285(97)00005-1}.
\bibitem[{Pan et~al.(2009)Pan, Chou and Tseng}]{Pan2009}
\bibinfo{author}{Pan, K.L.}, \bibinfo{author}{Chou, P.C.},
  \bibinfo{author}{Tseng, Y.J.}, \bibinfo{year}{2009}.
\newblock \bibinfo{title}{Binary droplet collision at high weber number}.
\newblock \bibinfo{journal}{Physical Review E} \bibinfo{volume}{80},
  \bibinfo{pages}{036301}.
\newblock \DOIprefix\doi{10.1103/PhysRevE.80.036301}.
\bibitem[{Pan et~al.(2019)Pan, Huang, Hsieh and Lu}]{Pan2019}
\bibinfo{author}{Pan, K.L.}, \bibinfo{author}{Huang, K.L.},
  \bibinfo{author}{Hsieh, W.T.}, \bibinfo{author}{Lu, C.R.},
  \bibinfo{year}{2019}.
\newblock \bibinfo{title}{Rotational separation after temporary coalescence in
  binary droplet collisions}.
\newblock \bibinfo{journal}{Phys. Rev. Fluids} \bibinfo{volume}{4},
  \bibinfo{pages}{123602}.
\newblock \DOIprefix\doi{10.1103/PhysRevFluids.4.123602}.
\bibitem[{Pan et~al.(2008)Pan, Law and Zhou}]{Pan2008}
\bibinfo{author}{Pan, K.L.}, \bibinfo{author}{Law, C.K.},
  \bibinfo{author}{Zhou, B.}, \bibinfo{year}{2008}.
\newblock \bibinfo{title}{Experimental and mechanistic description of merging
  and bouncing in head-on binary droplet collision}.
\newblock \bibinfo{journal}{Journal of Applied Physics} \bibinfo{volume}{103},
  \bibinfo{pages}{064901}.
\newblock \DOIprefix\doi{10.1063/1.2841055}.
\bibitem[{Pan and Suga(2005)}]{Pan2005}
\bibinfo{author}{Pan, Y.}, \bibinfo{author}{Suga, K.}, \bibinfo{year}{2005}.
\newblock \bibinfo{title}{Numerical simulation of binary liquid droplet
  collision}.
\newblock \bibinfo{journal}{Physics of Fluids} \bibinfo{volume}{17},
  \bibinfo{pages}{082105}.
\newblock \DOIprefix\doi{10.1063/1.2009527}.
\bibitem[{Patel et~al.(2017)Patel, Das, Kuipers, Padding and
  Peters}]{Patel2017}
\bibinfo{author}{Patel, H.V.}, \bibinfo{author}{Das, S.},
  \bibinfo{author}{Kuipers, J.A.M.}, \bibinfo{author}{Padding, J.T.},
  \bibinfo{author}{Peters, E.A.J.F.}, \bibinfo{year}{2017}.
\newblock \bibinfo{title}{A coupled volume of fluid and immersed boundary
  method for simulating 3d multiphase flows with contact line dynamics in
  complex geometries}.
\newblock \bibinfo{journal}{Chemical Engineering Science}
  \bibinfo{volume}{166}, \bibinfo{pages}{28--41}.
\newblock \DOIprefix\doi{10.1016/j.ces.2017.03.012}.
\bibitem[{Pathak and Raessi(2016)}]{Pathak2016}
\bibinfo{author}{Pathak, A.}, \bibinfo{author}{Raessi, M.},
  \bibinfo{year}{2016}.
\newblock \bibinfo{title}{A three-dimensional volume-of-fluid method for
  reconstructing and advecting three-material interfaces forming contact
  lines}.
\newblock \bibinfo{journal}{Journal of Computational Physics}
  \bibinfo{volume}{307}, \bibinfo{pages}{550--573}.
\newblock \DOIprefix\doi{10.1016/j.jcp.2015.11.062}.
\bibitem[{Planchette and Brenn(2009)}]{Planchette2009}
\bibinfo{author}{Planchette, C.}, \bibinfo{author}{Brenn, G.},
  \bibinfo{year}{2009}.
\newblock \bibinfo{title}{{Liquid encapsulation by binary collisions of
  immiscible liquid drops}}, in: \bibinfo{booktitle}{ICLASS 2009, 11th
  Triennial International Annual Conference on Liquid Atomization and Spray
  Systems}.
\bibitem[{Planchette et~al.(2017)Planchette, Hinterbichler, Liu, Bothe and
  Brenn}]{Planchette2017}
\bibinfo{author}{Planchette, C.}, \bibinfo{author}{Hinterbichler, H.},
  \bibinfo{author}{Liu, M.}, \bibinfo{author}{Bothe, D.},
  \bibinfo{author}{Brenn, G.}, \bibinfo{year}{2017}.
\newblock \bibinfo{title}{{Colliding drops as coalescing and fragmenting liquid
  springs}}.
\newblock \bibinfo{journal}{Journal of Fluid Mechanics} \bibinfo{volume}{814},
  \bibinfo{pages}{277--300}.
\newblock \DOIprefix\doi{10.1017/jfm.2016.852}.
\bibitem[{Planchette et~al.(2010)Planchette, Lorenceau and
  Brenn}]{Planchette2010}
\bibinfo{author}{Planchette, C.}, \bibinfo{author}{Lorenceau, E.},
  \bibinfo{author}{Brenn, G.}, \bibinfo{year}{2010}.
\newblock \bibinfo{title}{Liquid encapsulation by binary collisions of
  immiscible liquid drops}.
\newblock \bibinfo{journal}{Colloids and Surfaces A: Physicochemical and
  Engineering Aspects} \bibinfo{volume}{365}, \bibinfo{pages}{89--94}.
\newblock \DOIprefix\doi{10.1016/j.colsurfa.2009.12.011}. \bibinfo{note}{4th
  International Workshop}.
\bibitem[{Planchette et~al.(2011)Planchette, Lorenceau and
  Brenn}]{Planchette2011}
\bibinfo{author}{Planchette, C.}, \bibinfo{author}{Lorenceau, E.},
  \bibinfo{author}{Brenn, G.}, \bibinfo{year}{2011}.
\newblock \bibinfo{title}{{Binary collisions of immiscible liquid drops for
  liquid encapsulation}}.
\newblock \bibinfo{journal}{Fluid dyn. and mat. proc.} \bibinfo{volume}{7},
  \bibinfo{pages}{279--301}.
\newblock \DOIprefix\doi{10.3970/fdmp.2011.007.279}.
\bibitem[{Planchette et~al.(2012)Planchette, Lorenceau and
  Brenn}]{Planchette2012}
\bibinfo{author}{Planchette, C.}, \bibinfo{author}{Lorenceau, E.},
  \bibinfo{author}{Brenn, G.}, \bibinfo{year}{2012}.
\newblock \bibinfo{title}{{The onset of fragmentation in binary liquid drop
  collisions}}, in: \bibinfo{booktitle}{ICLASS 2012, 12th Triennial
  International Conference on Liquid Atomization and Spray Systems}.
\bibitem[{Popinet(2009)}]{Popinet2009}
\bibinfo{author}{Popinet, S.}, \bibinfo{year}{2009}.
\newblock \bibinfo{title}{An accurate adaptive solver for
  surface-tension-driven interfacial flows}.
\newblock \bibinfo{journal}{Journal of Computational Physics}
  \bibinfo{volume}{228}, \bibinfo{pages}{5838--5866}.
\newblock \DOIprefix\doi{10.1016/j.jcp.2009.04.042}.
\bibitem[{Potyka et~al.(2022)Potyka, Stober, Wurst, Ibach, Steigerwald, Weigand
  and Schulte}]{HLRSBericht2022Preprint}
\bibinfo{author}{Potyka, J.}, \bibinfo{author}{Stober, J.},
  \bibinfo{author}{Wurst, J.}, \bibinfo{author}{Ibach, M.},
  \bibinfo{author}{Steigerwald, J.}, \bibinfo{author}{Weigand, B.},
  \bibinfo{author}{Schulte, K.}, \bibinfo{year}{2022}.
\newblock \bibinfo{title}{{Towards DNS of Droplet-Jet Collisions of Immiscible
  Liquids with FS3D (accepted for proceedings High Performance Computing in
  Science and Engineering '22, arXiv preprint)}}.
\newblock \DOIprefix\doi{10.48550/ARXIV.2212.09727}.
\bibitem[{Qian and Law(1997)}]{Qian1997}
\bibinfo{author}{Qian, J.}, \bibinfo{author}{Law, C.K.}, \bibinfo{year}{1997}.
\newblock \bibinfo{title}{Regimes of coalescence and separation in droplet
  collision}.
\newblock \bibinfo{journal}{Journal of Fluid Mechanics} \bibinfo{volume}{331},
  \bibinfo{pages}{59--80}.
\newblock \DOIprefix\doi{10.1017/S0022112096003722}.
\bibitem[{Rider and Kothe(1998)}]{Rider1998}
\bibinfo{author}{Rider, W.J.}, \bibinfo{author}{Kothe, D.B.},
  \bibinfo{year}{1998}.
\newblock \bibinfo{title}{Reconstructing volume tracking}.
\newblock \bibinfo{journal}{Journal of Computational Physics}
  \bibinfo{volume}{141}, \bibinfo{pages}{112--152}.
\newblock \DOIprefix\doi{10.1006/jcph.1998.5906}.
\bibitem[{Rieber(2004)}]{Rieber2004}
\bibinfo{author}{Rieber, M.}, \bibinfo{year}{2004}.
\newblock \bibinfo{title}{{Numerische Modellierung der Dynamik freier
  Grenzfl\"achen in Zweiphasenstr\"omungen}}.
\newblock Ph.D. thesis. {University of Stuttgart}.
\bibitem[{Rieber and Frohn(1995)}]{Rieber1995}
\bibinfo{author}{Rieber, M.}, \bibinfo{author}{Frohn, A.},
  \bibinfo{year}{1995}.
\newblock \bibinfo{title}{Three-dimensional navier-stokes simulations of binary
  collision between droplets of equal size}.
\newblock \bibinfo{journal}{Journal of Aerosol Science} \bibinfo{volume}{26},
  \bibinfo{pages}{S929--S930}.
\newblock \DOIprefix\doi{10.1016/0021-8502(95)97372-L}.
\bibitem[{Roisman et~al.(2012)Roisman, Planchette, Lorenceau and
  Brenn}]{Roisman2012}
\bibinfo{author}{Roisman, I.V.}, \bibinfo{author}{Planchette, C.},
  \bibinfo{author}{Lorenceau, E.}, \bibinfo{author}{Brenn, G.},
  \bibinfo{year}{2012}.
\newblock \bibinfo{title}{Binary collisions of drops of immiscible liquids}.
\newblock \bibinfo{journal}{Journal of Fluid Mechanics} \bibinfo{volume}{690},
  \bibinfo{pages}{512--535}.
\newblock \DOIprefix\doi{10.1017/jfm.2011.459}.
\bibitem[{Roth et~al.(2007)Roth, Rabe, Weigand, Feuillebois and
  Malet}]{Roth2007}
\bibinfo{author}{Roth, N.}, \bibinfo{author}{Rabe, C.},
  \bibinfo{author}{Weigand, B.}, \bibinfo{author}{Feuillebois, F.},
  \bibinfo{author}{Malet, J.}, \bibinfo{year}{2007}.
\newblock \bibinfo{title}{{Droplet collision outcomes at high Weber number}},
  in: \bibinfo{booktitle}{ILASS -- 21st Annual Conference on Liquid Atomization
  and Spray Systems, Mugla, Turkey}.
\bibitem[{Roth et~al.(1999)Roth, Rieber and Frohn}]{Roth1999}
\bibinfo{author}{Roth, N.}, \bibinfo{author}{Rieber, M.},
  \bibinfo{author}{Frohn, A.}, \bibinfo{year}{1999}.
\newblock \bibinfo{title}{High energy head-on collision of droplets}, in:
  \bibinfo{editor}{Lavergne, G.} (Ed.), \bibinfo{booktitle}{ILASS 1999 Annual
  Conference on Liquid Atomization and Spray Systems, Toulouse, France}.
\bibitem[{Sakakibara and Inamuro(2008)}]{Sakakibara2008}
\bibinfo{author}{Sakakibara, B.}, \bibinfo{author}{Inamuro, T.},
  \bibinfo{year}{2008}.
\newblock \bibinfo{title}{Lattice boltzmann simulation of collision dynamics of
  two unequal-size droplets}.
\newblock \bibinfo{journal}{International Journal of Heat Mass Transfer}
  \bibinfo{volume}{51}, \bibinfo{pages}{3207--3216}.
\newblock \DOIprefix\doi{10.1016/j.ijheatmasstransfer.2008.02.004}.
\bibitem[{Schelkle and Frohn(1995)}]{Schelkle1995}
\bibinfo{author}{Schelkle, M.}, \bibinfo{author}{Frohn, A.},
  \bibinfo{year}{1995}.
\newblock \bibinfo{title}{Three-dimensional lattice boltzmann simulations of
  binary collision between equal droplets}.
\newblock \bibinfo{journal}{Journal of Aerosol Science} \bibinfo{volume}{26},
  \bibinfo{pages}{S145--S146}.
\newblock \DOIprefix\doi{10.1016/0021-8502(95)96980-L}.
\bibitem[{Shi et~al.(2016)Shi, Tang and Wang}]{Shi2016}
\bibinfo{author}{Shi, Y.}, \bibinfo{author}{Tang, G.}, \bibinfo{author}{Wang,
  Y.}, \bibinfo{year}{2016}.
\newblock \bibinfo{title}{{Simulation of three-component fluid flows using the
  multiphase lattice Boltzmann flux solver}}.
\newblock \bibinfo{journal}{Journal of Computational Physics}
  \bibinfo{volume}{314}, \bibinfo{pages}{228--243}.
\newblock \DOIprefix\doi{10.1016/j.jcp.2016.03.011}.
\bibitem[{Smith et~al.(2002)Smith, Solis and Chopp}]{Smith2002}
\bibinfo{author}{Smith, K.A.}, \bibinfo{author}{Solis, F.J.},
  \bibinfo{author}{Chopp, D.}, \bibinfo{year}{2002}.
\newblock \bibinfo{title}{A projection method for motion of triple junctions by
  level sets}.
\newblock \bibinfo{journal}{Interfaces Free Boundaries} \bibinfo{volume}{4},
  \bibinfo{pages}{263–276}.
\newblock \DOIprefix\doi{10.4171/IFB/61}.
\bibitem[{Sommerfeld and Pasternak(2019)}]{Sommerfeld2019}
\bibinfo{author}{Sommerfeld, M.}, \bibinfo{author}{Pasternak, L.},
  \bibinfo{year}{2019}.
\newblock \bibinfo{title}{Advances in modelling of binary droplet collision
  outcomes in sprays: A review of available knowledge}.
\newblock \bibinfo{journal}{International Journal of Multiphase Flow}
  \bibinfo{volume}{117}, \bibinfo{pages}{182--205}.
\newblock \DOIprefix\doi{10.1016/j.ijmultiphaseflow.2019.05.001}.
\bibitem[{Strang(1968)}]{Strang1968}
\bibinfo{author}{Strang, G.}, \bibinfo{year}{1968}.
\newblock \bibinfo{title}{{On the Construction and Comparison of Difference
  Schemes}}.
\newblock \bibinfo{journal}{SIAM Journal on Numerical Analysis}
  \bibinfo{volume}{5}, \bibinfo{pages}{506--517}.
\newblock \DOIprefix\doi{10.1137/0705041}.
\bibitem[{Suo and Jia(2020)}]{Suo2020}
\bibinfo{author}{Suo, S.}, \bibinfo{author}{Jia, M.}, \bibinfo{year}{2020}.
\newblock \bibinfo{title}{Correction and improvement of a widely used
  droplet–droplet collision outcome model}.
\newblock \bibinfo{journal}{Physics of Fluids} \bibinfo{volume}{32},
  \bibinfo{pages}{111705}.
\newblock \DOIprefix\doi{10.1063/5.0029463}.
\bibitem[{Sussman(2001)}]{Sussman2001}
\bibinfo{author}{Sussman, M.}, \bibinfo{year}{2001}.
\newblock \bibinfo{title}{{Adaptive Method of Lines}}.
  \bibinfo{publisher}{Chapman \& Hall/CRC}. chapter \bibinfo{chapter}{{An
  Adaptive Mesh Algorithm for Free Surface Flows in General Geometries}}.
\newblock pp. \bibinfo{pages}{232--258}.
\bibitem[{Suzuki et~al.(2021)Suzuki, Inamuro, Nakamura, Horai, Pan and
  Yoshino}]{Suzuki2021}
\bibinfo{author}{Suzuki, K.}, \bibinfo{author}{Inamuro, T.},
  \bibinfo{author}{Nakamura, A.}, \bibinfo{author}{Horai, F.},
  \bibinfo{author}{Pan, K.L.}, \bibinfo{author}{Yoshino, M.},
  \bibinfo{year}{2021}.
\newblock \bibinfo{title}{{Simple extended lattice Boltzmann methods for
  incompressible viscous single-phase and two-phase fluid flows}}.
\newblock \bibinfo{journal}{Physics of Fluids} \bibinfo{volume}{33},
  \bibinfo{pages}{037118}.
\newblock \DOIprefix\doi{10.1063/5.0041854}.
\bibitem[{Tsuru et~al.(2010)Tsuru, Tajima, Ishibashi and Kawauchi}]{Tsuru2010}
\bibinfo{author}{Tsuru, D.}, \bibinfo{author}{Tajima, H.},
  \bibinfo{author}{Ishibashi, R.}, \bibinfo{author}{Kawauchi, S.},
  \bibinfo{year}{2010}.
\newblock \bibinfo{title}{{Droplet Collision Modelling between Merging
  Immiscible Sprays in Direct Water Injection System}}, in:
  \bibinfo{booktitle}{ILASS -- Europe 2010, 23rd Annual Conference on Liquid
  Atomization and Spray Systems}.
\bibitem[{Visser et~al.(2018)Visser, Kamperman, Karbaat, Lohse and
  Karperien}]{Visser2018}
\bibinfo{author}{Visser, C.W.}, \bibinfo{author}{Kamperman, T.},
  \bibinfo{author}{Karbaat, L.P.}, \bibinfo{author}{Lohse, D.},
  \bibinfo{author}{Karperien, M.}, \bibinfo{year}{2018}.
\newblock \bibinfo{title}{In-air microfluidics enables rapid fabrication of
  emulsions, suspensions, and {3D} modular (bio)materials}.
\newblock \bibinfo{journal}{Science Advances} \bibinfo{volume}{4},
  \bibinfo{pages}{eaao1175}.
\newblock \DOIprefix\doi{10.1126/sciadv.aao1175}.
\bibitem[{Wang et~al.(2004)Wang, Lin, Hung, Huang and Law}]{Wang2004}
\bibinfo{author}{Wang, C.H.}, \bibinfo{author}{Lin, C.Z.},
  \bibinfo{author}{Hung, W.G.}, \bibinfo{author}{Huang, W.C.},
  \bibinfo{author}{Law, C.K.}, \bibinfo{year}{2004}.
\newblock \bibinfo{title}{On the burning characteristics of collision-generated
  water/hexadecane droplets}.
\newblock \bibinfo{journal}{Combustion Science and Technology}
  \bibinfo{volume}{176}, \bibinfo{pages}{71--93}.
\newblock \DOIprefix\doi{10.1080/00102200490255361}.
\bibitem[{Washino et~al.(2010)Washino, Tan, Salman and Hounslow}]{Washino2010}
\bibinfo{author}{Washino, K.}, \bibinfo{author}{Tan, H.},
  \bibinfo{author}{Salman, A.}, \bibinfo{author}{Hounslow, M.},
  \bibinfo{year}{2010}.
\newblock \bibinfo{title}{{Direct numerical simulation of solid-liquid-gas
  three-phase flow: Fluid-solid interaction}}.
\newblock \bibinfo{journal}{Powder Technology} \bibinfo{volume}{206},
  \bibinfo{pages}{161--169}.
\newblock \DOIprefix\doi{10.1016/j.powtec.2010.07.015}.
\bibitem[{Wauligmann et~al.(2021)Wauligmann, D\"urrw\"achter, Offenh\"auser,
  Schlottke, Bernreuther and Dick}]{Wauligmann2021}
\bibinfo{author}{Wauligmann, P.}, \bibinfo{author}{D\"urrw\"achter, J.},
  \bibinfo{author}{Offenh\"auser, P.}, \bibinfo{author}{Schlottke, A.},
  \bibinfo{author}{Bernreuther, M.}, \bibinfo{author}{Dick, B.},
  \bibinfo{year}{2021}.
\newblock \bibinfo{title}{Node-level performance optimizations in cfd codes},
  in: \bibinfo{booktitle}{The International Conference on High Performance
  Computing in Asia-Pacific Region Companion}, pp. \bibinfo{pages}{7--8}.
\bibitem[{Weigand et~al.(2021)Weigand, Schulte and Tropea}]{ICLASS2021}
\bibinfo{author}{Weigand, B.}, \bibinfo{author}{Schulte, K.},
  \bibinfo{author}{Tropea, C.}, \bibinfo{year}{2021}.
\newblock \bibinfo{title}{{Selected Results of the Collaborative Research
  Center Droplet Dynamics under Extreme Ambient Conditions SFB-TRR 75}}, in:
  \bibinfo{booktitle}{ICLASS Edinburgh}.
\newblock \DOIprefix\doi{10.2218/iclass.2021.5805}. \bibinfo{note}{paper 27}.
\bibitem[{White(1991)}]{White1991}
\bibinfo{author}{White, F.M.}, \bibinfo{year}{1991}.
\newblock \bibinfo{title}{{Viscous Fluid Flow}}.
  \bibinfo{publisher}{McGraw-Hill, Inc.}
\newblock p.~\bibinfo{pages}{72}.
\bibitem[{Willis and Orme(2000)}]{Willis2000}
\bibinfo{author}{Willis, K.}, \bibinfo{author}{Orme, M.}, \bibinfo{year}{2000}.
\newblock \bibinfo{title}{Experiments on the dynamics of droplet collisions in
  a vacuum}.
\newblock \bibinfo{journal}{Experiments in Fluids} \bibinfo{volume}{29},
  \bibinfo{pages}{347--358}.
\newblock \DOIprefix\doi{10.1007/s003489900092}.
\bibitem[{Willis and Orme(2003)}]{Willis2003}
\bibinfo{author}{Willis, K.}, \bibinfo{author}{Orme, M.}, \bibinfo{year}{2003}.
\newblock \bibinfo{title}{Binary droplet collisions in a vacuum environment: An
  experimental investigation of the role of viscosity}.
\newblock \bibinfo{journal}{Experiments in Fluids} \bibinfo{volume}{34},
  \bibinfo{pages}{28 -- 41}.
\newblock \DOIprefix\doi{10.1007/s00348-002-0526-4}.
\bibitem[{W\"ohrwag et~al.(2018)W\"ohrwag, Semprebon, Mazloomi~Moqaddam, Karlin
  and Kusumaatmaja}]{Woehrwag2018}
\bibinfo{author}{W\"ohrwag, M.}, \bibinfo{author}{Semprebon, C.},
  \bibinfo{author}{Mazloomi~Moqaddam, A.}, \bibinfo{author}{Karlin, I.},
  \bibinfo{author}{Kusumaatmaja, H.}, \bibinfo{year}{2018}.
\newblock \bibinfo{title}{Ternary free-energy entropic lattice boltzmann model
  with a high density ratio}.
\newblock \bibinfo{journal}{Physical Review Letters} \bibinfo{volume}{120},
  \bibinfo{pages}{234501}.
\newblock \DOIprefix\doi{10.1103/PhysRevLett.120.234501}.
\bibitem[{Youngs(1982)}]{Youngs1982}
\bibinfo{author}{Youngs, D.}, \bibinfo{year}{1982}.
\newblock \bibinfo{title}{Numerical Methods in Fluid Dynamics}.
  volume~\bibinfo{volume}{24}. chapter \bibinfo{chapter}{Time-Dependent
  Multi-material Flow with Large Fluid Distortion}.
\newblock pp. \bibinfo{pages}{273--285}.
\bibitem[{Zhang and Menshov(2019)}]{Zhang2019}
\bibinfo{author}{Zhang, C.}, \bibinfo{author}{Menshov, I.},
  \bibinfo{year}{2019}.
\newblock \bibinfo{title}{Eulerian modelling of compressible three-fluid flows
  with surface tension}.
\newblock \bibinfo{journal}{Russian Journal of Numerical Analysis and
  Mathematical Modelling} \bibinfo{volume}{34}, \bibinfo{pages}{225--240}.
\newblock \DOIprefix\doi{10.1515/rnam-2019-0019}.
\bibitem[{Zhang et~al.(2020)Zhang, Liu and Ding}]{Zhang2020}
\bibinfo{author}{Zhang, J.T.}, \bibinfo{author}{Liu, H.R.},
  \bibinfo{author}{Ding, H.}, \bibinfo{year}{2020}.
\newblock \bibinfo{title}{{Head-on collision of two immiscible droplets of
  different components}}.
\newblock \bibinfo{journal}{Physics of Fluids} \bibinfo{volume}{32},
  \bibinfo{pages}{082106}.
\newblock \DOIprefix\doi{10.1063/5.0018391}.

\end{thebibliography}
